\documentclass[pra,twocolumn,superscriptaddress,a4paper,longbibliography]{revtex4-1}
\usepackage{graphicx}
\usepackage{amssymb,amsmath}
\usepackage{algorithm}
\usepackage[noend]{algpseudocode}
\usepackage{optidef}
\usepackage{color}
\usepackage{epstopdf}
\definecolor{ao(english)}{rgb}{0.0, 0.5, 0.0}
\usepackage[colorlinks=true,citecolor=blue,linkcolor=blue,urlcolor=blue,breaklinks=talse]{hyperref}

\begin{document}
\title{Quantum process tomography via completely positive and trace-preserving projection}
\author{George~C.~Knee}
\email[email: ]{gk@physics.org}
\affiliation{Department of Physics, University of Warwick, Coventry, CV4 7AL, UK}
\author{Eliot~Bolduc}
\affiliation{SUPA, Institute of Photonics and Quantum Sciences, Heriot-Watt University, David Brewster Building, Edinburgh, EH14 4AS, UK}
\author{Jonathan~Leach}
\affiliation{SUPA, Institute of Photonics and Quantum Sciences, Heriot-Watt University, David Brewster Building, Edinburgh, EH14 4AS, UK}
\author{Erik~M.~Gauger}
\affiliation{SUPA, Institute of Photonics and Quantum Sciences, Heriot-Watt University, David Brewster Building, Edinburgh, EH14 4AS, UK}

\date{\today}                                           

\begin{abstract}
We present an algorithm for projecting superoperators onto the set of completely positive, trace-preserving maps. When combined with gradient descent of a cost function, the procedure results in an algorithm for quantum process tomography: finding the quantum process that best fits a set of sufficient observations. We compare the performance of our algorithm to the diluted iterative algorithm as well as second-order solvers interfaced with the popular {\sc cvx} package for {\sc matlab}, and find it to be significantly faster and more accurate while guaranteeing a physical estimate. 
\end{abstract}
\maketitle
\section{Introduction}
\begin{figure}[h!]
\includegraphics[width=\columnwidth]{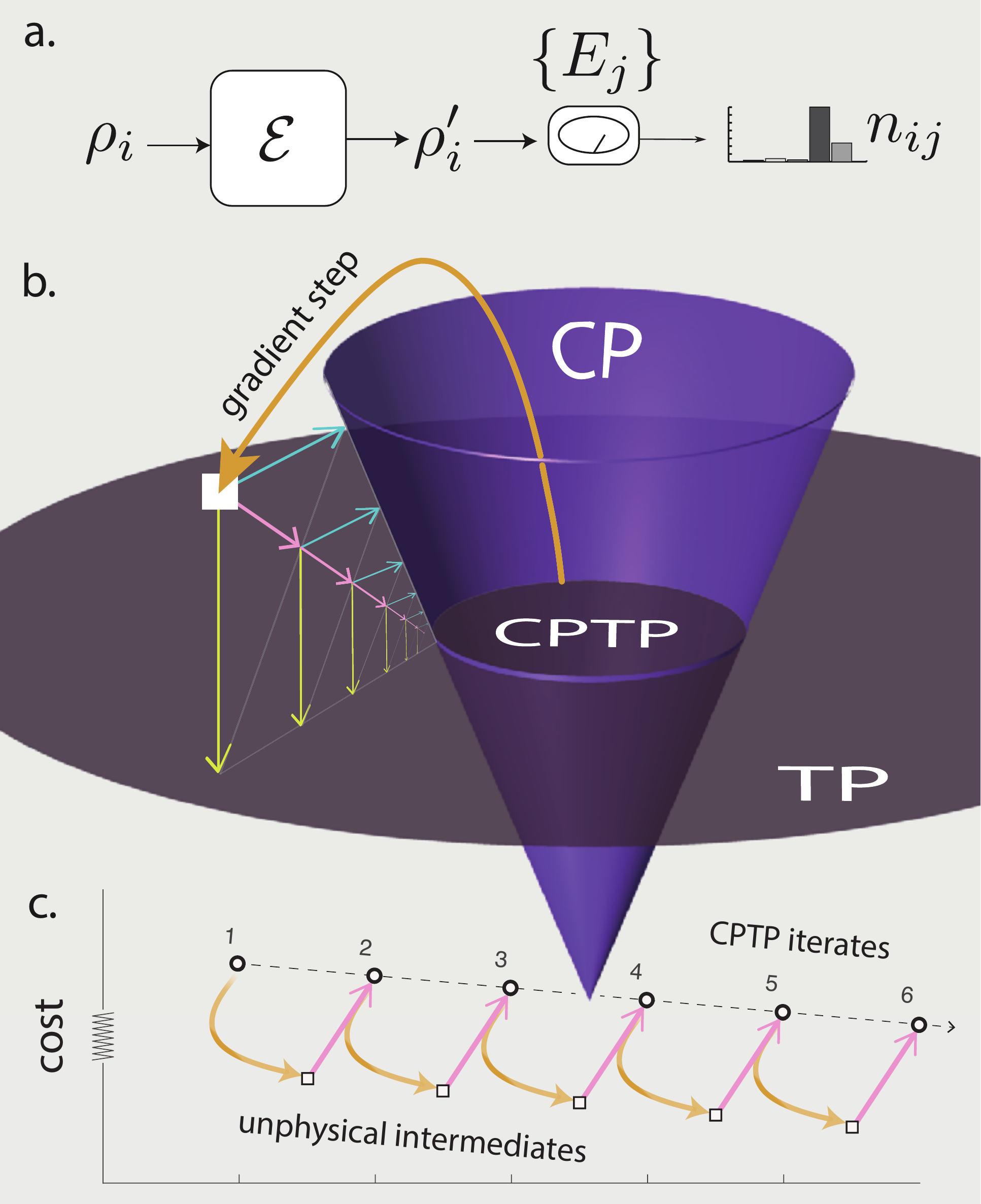}
\caption{\label{cone}\textbf{a.} An unknown process $\mathcal{E}$ maps a known input state $\rho_i$ into an unknown output state $\rho'_i$. When measured with a set of positive operators $\{E_j\}$, $\rho'_i$ produces a measurement histogram $n_{ij}$, leading to a cost (e.g. negative likelihood) for each possible estimate for $\mathcal{E}$. \textbf{b.} Finding the best fit to the data is a constrained optimisation problem that we solve with projected gradient descent. The desired solution set is  CPTP: the set of completely positive and trace preserving maps (the intersection of a cone with a hyperplane). Gradient descent can cause an iterate to exit this set, but it is repeatedly projected back. The schematic shows a single iteration of the algorithm: an unphysical candidate ($\Box$) is iteratively moved to e.g. the average (pink) of the projection onto CP (cyan) and TP (yellow arrows). \textbf{c.} The overall procedure reduces the cost function while ensuring iterates are in CPTP. }
\end{figure}

As experimental quantum information science progresses, researchers are increasingly turning to the characterisation of quantum \emph{processes} (see Fig~\ref{cone}a) rather than the quantum \emph{states} which they act upon. For example, the fault-tolerance theorem~\cite{AharonovBen-Or1997} (which underpins the feasibility of large scale quantum computers) places a requirement on the accuracy of operations performed on quantum bits. In fact, the theorem relies on a worst-case analysis with respect to state preparation (before the operation) and with respect to measurement (after the operation) in judging their accuracy. Fortunately, to evaluate such metrics one does not need to actually prepare and measure the worst case: A complete description of the process can be inferred from a suitable set of preparations and measurements and subsequently any property of interest may then be calculated. Such inference is known as quantum process tomography (QPT).  QPT has been performed in many different physical settings~\cite{OBrienPrydeGilchrist2004,PachonMarcusAspuru-Guzik2015,BialczakAnsmannHofheinz2010,HowardTwamleyWittmann2006}: the challenge is producing an estimate of the process in a reasonable amount of time which matches the data as closely as possible and is consistent with a quantum mechanical model of the experiment. We present algorithms for reconstructing the process subject to appropriate constraints, via a subroutine that implements a composite projection onto the set of quantum channels-- see Fig~\ref{cone}b.

In Sec.~\ref{qpt}, we describe the maximum likelihood approach to process tomography, along with the Choi representation of processes and individual projections onto important constraint sets. In Sec.~\ref{cptppgdblifp} we describe our main results, including the composite projection onto quantum channels, and two algorithms that exploit it to estimate a quantum process from data. Our algorithms are benchmarked against existing approaches in Sec.~\ref{benchmarking}, and conclusions are drawn in Sec~\ref{conclusions}. Further details are given in Appendices.

\section{Process tomography}
\label{qpt}
A quantum process is a linear `superoperator' which acts on an input density operator $\rho$ to produce an output density operator $\rho'$:
\begin{align}
\mathcal{E}(\rho)=\rho'.
\end{align}
Here $\rho$ and $\rho'$ are order 2 tensors (i.e. operators), which we assume (for simplicity) to act on Hilbert spaces of equal dimension $d$. $\mathcal{E}$ is an order 4 tensor specified by $d^4-d^2$ parameters, where $d$ is itself exponential in the number of subsystems. Assumption-free QPT is therefore necessarily very expensive: cheaper alternatives are made possible by relying on prior knowledge (e.g. compressive sensing~\cite{ShabaniKosutMohseni2011}) or by estimating a single summary statistic (as in randomised benchmarking~\cite{WallmanFlammia2014}), but such alternatives may be insufficient to decide whether $\mathcal{E}$  meets fault tolerance requirements~\cite{SandersWallmanSanders2016,Blume-KohoutGambleNielsen2017}. It is therefore important to reduce the costs associated with assumption-free QPT while maintaining accuracy.

\subsection{Estimation of processes}The probability of observing an outcome corresponding to $E_j$ (a positive measurement operator), when the quantum process has transformed some input state $\rho_i$, is
\begin{align}
p_{ij} = \textrm{Tr}( \mathcal{E}(\rho_i)E_j).
\end{align}
For each input state, $N_i$ samples are drawn and the number of outcomes $n_{ij}$ that correspond to each outcome $j$ are recorded so that $\sum_j n_{ij}=N_i$.  	In order to produce an estimate for a quantum process from such a set of observations, a cost function (or measure of `un-fitness') $f(\mathcal{E})$ is chosen and then minimised over a set of candidate processes. The minimisation is handled by, for example, gradient descent steps of the cost function. Maximum-likelihood (ML) estimation is a principled and prominent choice, and one that we make here in order to showcase our general method. The likelihood of $\mathcal{E}$ given the `data' $n_{ij}$ is
$
\mathcal{L} = \prod_{ij }p_{ij}^{n_{ij}}
$
up to an irrelevant constant. According to the principle of ML, we seek $\mathcal{E}^*$, the quantum process that maximises $\mathcal{L}$. This approach is justified heuristically by considering $\mathcal{L}$ as a measure of agreement between the model $p_{ij}$ and the data $n_{ij}$, assuming a multinomial statistical model. ML enjoys the key property of asymptotic efficiency: as the number of trials $N_i$ becomes large, under mild conditions~\cite{Daniels1961} the variance of the ML estimator is lower than any unbiased estimator, saturating the Cram\'er-Rao bound~\cite{KokLovett2010}. Alternative estimators may be preferable for finite $N_i$~\cite{SugiyamaTurnerMurao2013}, but will generally perform worse asymptotically in terms of their precision and accuracy. In the limit where $n_{ij}$ are all large, the likelihood may be approximated by a Gaussian function and the problem reduces to minimising the cost function $f(\mathcal{E})\rightarrow\sum_{ij}(p_{ij}-n_{ij})^2$~\cite{SmolinGambettaSmith2012}, which may be solved by linear inversion~\cite{Teo2015}. This is an unjustified simplification in general, but may be transformed into an effective heuristic, as we will show below.  Regardless of the choice of cost function, our approach is flexible in that it is not tied (as some other methods are~\cite{SmolinGambettaSmith2012,ChuangNielsen1997}) to a particular choice of (preparation and/or) measurement operators. Proceeding with the (convex) cost function
\begin{align}
f(\mathcal{E})&=-\ln\mathcal{L}=-\sum_{ij} n_{ij} \ln p_{ij},
\label{KL}
\end{align}
we will introduce below an algorithm minimising $f$ subject to appropriate constraints. The minimiser of $f$ is also the maximiser of $\mathcal{L}$ due to the monotonicity of the logarithm.  Unconstrained tomographic procedures usually produce unphysical quantum states or processes~\cite{Teo2015}, often due to an implicit and idealised assumption of zero noise. A physical estimate is paramount if one is to use it in any further theoretical analysis, such as calculating purity, fidelity, entanglement, expectation values and so on. We will now introduce the constraints on $\mathcal{E}$ and our approach to enforcing them, which uses a certain representation of $\mathcal{E}$.
\subsection{Choi representation}
It is well known that quantum processes, considered as superoperators with dimension $d\times d \times d \times d$ may be represented as $d^2\times d^2$-dimensional operators on a Hilbert space $\mathcal{H}_{\textrm{in}}\otimes \mathcal{H}_{\textrm{out}}$. These are known as Choi operators, defined as:
\begin{align}
C_{\mathcal{E}} = \sum_{ij} |i\rangle\langle j  | \otimes  \mathcal{E}(|i\rangle\langle j  |)
\label{choidef}
\end{align}
for an orthonormal set of kets $\{|i\rangle \}$ which form a basis of both the input and output Hilbert spaces. This relation implies that $\mathcal{E}$ may be inferred by applying it to one part of a $d^2$-dimensional system prepared in a \emph{single} maximally entangled state, and performing quantum state tomography on the output: a technique known as ancilla-assisted QPT~\cite{Leung2003,AltepeterBranningJeffrey2003}.  Here we continue with the standard approach which uses multiple input states in $d$ dimensions, and considers the complete dataset as a whole (rather than simply performing state tomography on each output state independently~\cite{ChuangNielsen1997}). One can easily verify that $\mathcal{E}(\rho)=\textrm{Tr}_{\mathcal{H}_{\textrm{in}}}[(\rho^T\otimes \mathbb{I})C_{\mathcal{E}}]$ which implies that the forward model can be rewritten as
\begin{align}
p_{ij} = \textrm{Tr}([ \rho_i^T  \otimes E_j]C_{\mathcal{E}}).
\label{forward_model}
\end{align}
Now the cost function has a new argument $f(\mathcal{E})\rightarrow f(C_{\mathcal{E}})$. The gradient of $f$ with respect to this new argument follows from operator calculus 
\begin{align}
\nabla f (C_{\mathcal{E}}) =\frac{\partial  f (C_{\mathcal{E}})}{\partial C_{\mathcal{E}}}= -\sum_{ij} \frac{n_{ij}}{p_{ij}} \rho_i \otimes E_j^T.
\end{align}
These expressions can be vectorised to enable faster evaluation on a computer -- see Appendix~\ref{app:vector}.
\subsection{Constraints and projections}
Just as some $d\times d$ operators fail to qualify as proper density matrices (quantum states), $\mathcal{E}$ must satisfy certain constraints in order for it to represent a proper quantum process. Firstly, it should be completely positive (CP). This means that it must preserve the positive semidefiniteness of an arbitrary input state when acting on only a part of a larger space of arbitrary dimension: $\rho\succeq 0\Rightarrow[\mathcal{E}\otimes \mathbb{I}_{d'}](\rho)\succeq 0 \quad \forall d'$.  A theorem due to Choi~\cite{Choi1975} states that the complete positivity of $\mathcal{E}$ is equivalent to the positive semidefiniteness of $C_{\mathcal{E}}$~\footnote{A positive semidefinite matrix is denoted by the expression $C\succeq 0$, meaning that $C$ belongs to the subset of Hermitian matrices with nonnegative eigenvalues.}:
\begin{align}
\mathcal{E}\in \textrm{CP} \Leftrightarrow C_{\mathcal{E}}\succeq0.
\end{align}
Since CP is a convex set, we can find the closest CP map to any given Hermitian superoperator by using a projection of its Choi matrix onto the set of positive semidefinite operators. This is achieved via
\begin{align}
\mathcal{CP}(\mathcal{C}_{\mathcal{E}})=V\max(D,0)\, V^\dagger
\end{align}
with $\mathcal{C}_{\mathcal{E}}=VDV^\dagger$: i.e. through removing negative terms in the eigendecomposition of $\mathcal{C}_{\mathcal{E}}$. $\mathcal{CP}$ is then a projection orthogonal under the Frobenius norm~\cite{BoydVandenberghe2009}.
Secondly, it is common to assume the process satisfies the property of trace preservation (TP): $\textrm{Tr}(\mathcal{E}(\rho))=\textrm{Tr}(\rho)\quad \forall \rho$. In the Choi operator picture, this is equivalent to the condition 
\begin{align}
\textrm{Tr}_{\mathcal{H}_{\textrm{out}}}(C_{\mathcal{E}})=\mathbb{I},\
\label{TPC}
\end{align}
and the TP set is therefore defined by an affine constraint~\cite{Plesnik2007}. In Appendices~~\ref{app:tp} and~\ref{app:tni}, we (respectively) explain why the TP constraint makes QPT a non-trivial extension of state tomography, and discuss relaxing the constraint to trace non-increasing (probabilistic) processes. We recast Eq.~\eqref{TPC} as $M\texttt{vec}[C_\mathcal{E}] = \texttt{vec}[\mathbb{I}]$, defining
\begin{align}
M = \sum_k \mathbb{I}\otimes \langle k | \otimes \mathbb{I} \otimes \langle k|
\end{align}
The {\tt vec}$[\cdot]$ operation reshapes its argument into a vector, and we denote the inverse operation by {\tt vec}$^{-1}[\cdot]$. 
The orthogonal projection onto TP is the solution to
\begin{align}
\min_{C} \frac{1}{2} ||\texttt{vec}[C]-\texttt{vec}[C_{\mathcal{E}}]||^2 \quad \textrm{s.t.} \quad M\texttt{vec}[C] = \texttt{vec}[\mathbb{I}]
\label{tpop}
\end{align}
given by
\begin{widetext}
\begin{align}
\mathcal{TP}(C_{\mathcal{E}}) &=\texttt{vec}^{-1}[\texttt{vec}[C_{\mathcal{E}}]- M^\dagger(MM^\dagger)^{-1}(M\texttt{vec}[C_{\mathcal{E}}]-\texttt{vec}[\mathbb{I}])]\nonumber\\
&=\texttt{vec}^{-1}[\texttt{vec}[C_{\mathcal{E}}]- \frac{1}{d} M^\dagger M \texttt{vec}[C_{\mathcal{E}}] +\frac{1}{d}M^\dagger \texttt{vec}[\mathbb{I}]]
.
\label{tp_projection}
\end{align}
\end{widetext}
In the second line, we used $MM^\dagger=d \mathbb{I}\otimes\mathbb{I}$; it is useful to note that $M^\dagger M = \sum_{ij} \mathbb{I}\otimes |j\rangle\langle i|\otimes \mathbb{I}\otimes|j\rangle\langle i|$. 
\section{Composite projection}
\label{cptppgdblifp}
It is not obvious that the ability to separately project into the CP and TP sets of superoperators enables projection into the intersection set CPTP.  While $\mathcal{CP}$ and $\mathcal{TP}$ are separately projections, their sequential action will not generally result in a matrix belonging to CPTP (unlike the equivalent sequential actions for projection onto quantum states~\cite{BolducKneeGauger2017}). This may be intuited with the help of the geometrical picture in Fig.~\ref{cone}b.
Repeated averaged projections, however, defined (at iteration $k$ of the outer loop of our algorithm) by
\begin{align}
H^0 = C^k_{\mathcal{E}}; \qquad H^{l+1} = \frac{1}{2}\left(\mathcal{TP}(H^l)+\mathcal{CP}(H^l)\right)
\end{align}
will converge (in linear time~\cite{EscalanteRaydan2011}) to a point in CPTP -- see Fig.~\ref{cone}b. Dykstra's alternating projection algorithm is a superior alternative, which we choose because it achieves projection $\mathcal{CPTP}(C_{\mathcal{E}})$ onto the closest point in the intersection of the sets~\cite{Birgin2009}. A similar method was used in Ref~\cite{DrusvyatskiyLiPelejo2015} to solve a feasibility problem where input and output states are given and an exact but non-unique solution is sought. By contrast the ML approach (which we adopt) finds the best possible fit to some given, noisy data and treats the more realistic situation where an exact solution cannot be expected. Pseudo-code for this subroutine (showing our use of a robust stopping criterion due to Birgin and Raydan~\cite{Birgin2009}) is given below. 
\begin{algorithm}[H]
\caption{$\mathcal{CPTP}$ projection subroutine }\label{CPTP}
\begin{algorithmic}[1]
\State{Input: $C$}
\State{Output: $C_\perp\in \text{CPTP}$ s.t. $||C-C_\perp||_2^2\leq||C-B||_2^2 \quad \forall B\in \text{CPTP}$}
\State{Set: ${p}_0={q}_0={y}_0=0, k=0, {x}_0 = \texttt{vec}[C]$}
\While{$||{p}_{k-1}-{p}_{k}||^2+||{q}_{k-1}-{q}_{k}||^2+|2{p}_{k-1}^\dagger({x}_{k}-{x}_{k-1})|+|2{q}_{k-1}^\dagger({y}_{k}-{y}_{k-1})|> 10^{-4}$}
\State{${y}_{k\phantom{+1}}=\texttt{vec}[\mathcal{TP}[\texttt{vec}^{-1}[{x}_k+{p}_k]]$}]
\State{${p}_{k+1} = {x}_k+{p}_k-{y}_k$\;}
\State{${x}_{k+1}=\texttt{vec}[\mathcal{CP}[\texttt{vec}^{-1}[{y}_k+{q}_k]]$}]
\State{${q}_{k+1} = {y}_k+{q}_k-{x}_{k+1}$\;}
\State{$k=k+1;$}
\EndWhile
\State{\textbf{Return} $C_\perp= \texttt{vec}^{-1}[{x}_{k+1}]$}
\end{algorithmic}
\end{algorithm}
\subsection{Proposed algorithm: pgdB}
Returning to the main problem of QPT, we are now ready to apply the principle of projected gradient descent. Such an approach has recently been shown to offer speed benefits when applied to quantum state tomography~\cite{GoncalvesGomes-RuggieroLavor2016,BolducKneeGauger2017,ShangZhangNg2017}. Applying it to QPT implies taking a single gradient descent step before running an alternating projection~\cite{HenrionMalick2012} under $\mathcal{CP}$ and $\mathcal{TP}$ until this `inner loop' converges, before taking another gradient descent step and repeating. Formally the ML-QPT problem is written
\begin{mini}
{C_{\mathcal{E}}\in \mathbb{C}^{d^2\times d^2}}{f=-\sum_{ij} n_{ij} \ln( \textrm{Tr}([ \rho_i^T  \otimes E_j]C_{\mathcal{E}}) )}
{\label{optiprob}}{}
\addConstraint{C_{\mathcal{E}}}{\succeq 0}{\textrm{\quad(CP constraint)}}
\addConstraint{\text{Tr}_{\mathcal{H}_{\textrm{out}}}(C_{\mathcal{E}})}{=\mathbb{I}_{d\times d}}{\textrm{\quad(TP constraint)}},
\end{mini}
which our proposed algorithm solves via the iterative update rule:
\begin{align}
C^0_{\mathcal{E}}&=\mathbb{I}_{d^2\times d^2}/d;\nonumber\\
\qquad C_{\mathcal{E}}^{k+1}&=\mathcal{CPTP}(C_{\mathcal{E}}^{k}-\frac{1}{\mu} \nabla f(C_{\mathcal{E}}^{k})).
\end{align}
Here $\mu$ is a step-size metaparameter. At each iteration a step is taken in the locally downhill direction of $f$, with the aim of increasing the likelihood, before a projection ensures that the constraints are satisfied (up to any desired tolerance) -- see Fig \ref{cone}c. Pseudocode of the full algorithm {\sc pgdb} is given below, and shows our implementation of backtracking (Armijo line search) to improve convergence times~\cite{BoydVandenberghe2009,Armijo1966,Bertsekas1976}.

\begin{algorithm}[H]
\caption{{ pgdB} }\label{pgdB}
\begin{algorithmic}[1]
\State{$k= 0$}
\State{Initial estimate: $C_{\mathcal{E}}^0 = \mathbb{I}_{d^2\times d^2}/d$}
\State{Set metaparameters: $\mu = 3/(2d^2), \gamma = 0.3$}
\While{$f(C_{\mathcal{E}}^k)-f(C_\mathcal{E}^{k+1})>1\times10^{-10}$ }
\State{$D^k=\mathcal{CPTP}[C_{\mathcal{E}}^k- \frac{1}{\mu} \nabla f(C_{\mathcal{E}}^k)]-C_{\mathcal{E}}^k$\;}
\State{$\alpha=1$}
\While{$f(C_{\mathcal{E}}^k+\alpha D^k)>f(C_{\mathcal{E}}^k)+\gamma\alpha \langle D^k ,\nabla f(C_{\mathcal{E}}^k)\rangle$}
\State{$\alpha=0.5\alpha$}
\EndWhile
\State{$C_{\mathcal{E}}^{k+1}=C_{\mathcal{E}}^k+\alpha D^k$}
\State{$k = k+1$}
\EndWhile
\State{\textbf{Return} $C_{\mathcal{E}_{\textrm{est}}}= C^{k+1}_{\mathcal{E}}$}
\end{algorithmic}
\end{algorithm}

It is important to point out that, although by our choice of $f$ the considered problem is convex (meaning there are no local optima which are not global optima) there may nevertheless be regions (where $p_{ij}=0$ for some $i,j$) where $f$ has an undefined gradient and where the likelihood is stationary~\cite{LiCevher2017}. These subspaces correspond to those processes which take $\rho_i$ to a state orthogonal to $E_j$. The probability of a randomly chosen process lying within these sub-dimensional regions is low. However, an iteration of our algorithm might encounter such a region, causing the algorithm to stall: this is especially the case since projections tend to result in rank deficient matrices. Our solution to this issue is presented in Appendix~\ref{stalling}. 

Note that our approach can be adapted to solve other constrained optimisation problems over quantum processes: whenever the cost function $f$ is convex and continuously differentiable the projected gradient descent procedure is guaranteed to converge to an optimal solution~\cite{Beck2014}.
\subsection{Linear inversion with a final projection (LIFP)} To show that our CPTP projection has applications in QPT other than for ML, we present a further algorithm based on linear inversion of a vectorised form of Eq.~\eqref{forward_model}, $\texttt{vec}(p_{ij}) =A\texttt{vec}(C_{\mathcal{E}})$, where the rows of $A$ are composed of $\texttt{vec}[\rho_i\otimes E_j^T]^T$. To allow for a comparison with the manifestly physical approach of the ML methods, we upgrade na\"ive linear inversion so as to force it to also produce physical estimates --- simply by applying our novel, composite CPTP projection just once. We call the resultant estimator Linear Inversion with a Final Projection:
\begin{align}
C^{LIFP}_\mathcal{E}:= \mathcal{CPTP}( {\tt vec}^{-1} [A^{+}n]),
\end{align}
where $A^+=A^\dagger(AA^\dagger)^{-1}$ is the pseudo-inverse of $A$~\cite{Teo2015} and $n=\texttt{vec}[n_{ij}]$. This approach is in the same spirit as the `quick and dirty' state tomography approach used by Kaznady and James~\cite{KaznadyJames2009}, suggested as an alternative to full-blown optimization. For full details, see Appendix~\ref{app:LIFP}.
\begin{figure*}
\includegraphics[width=\textwidth]{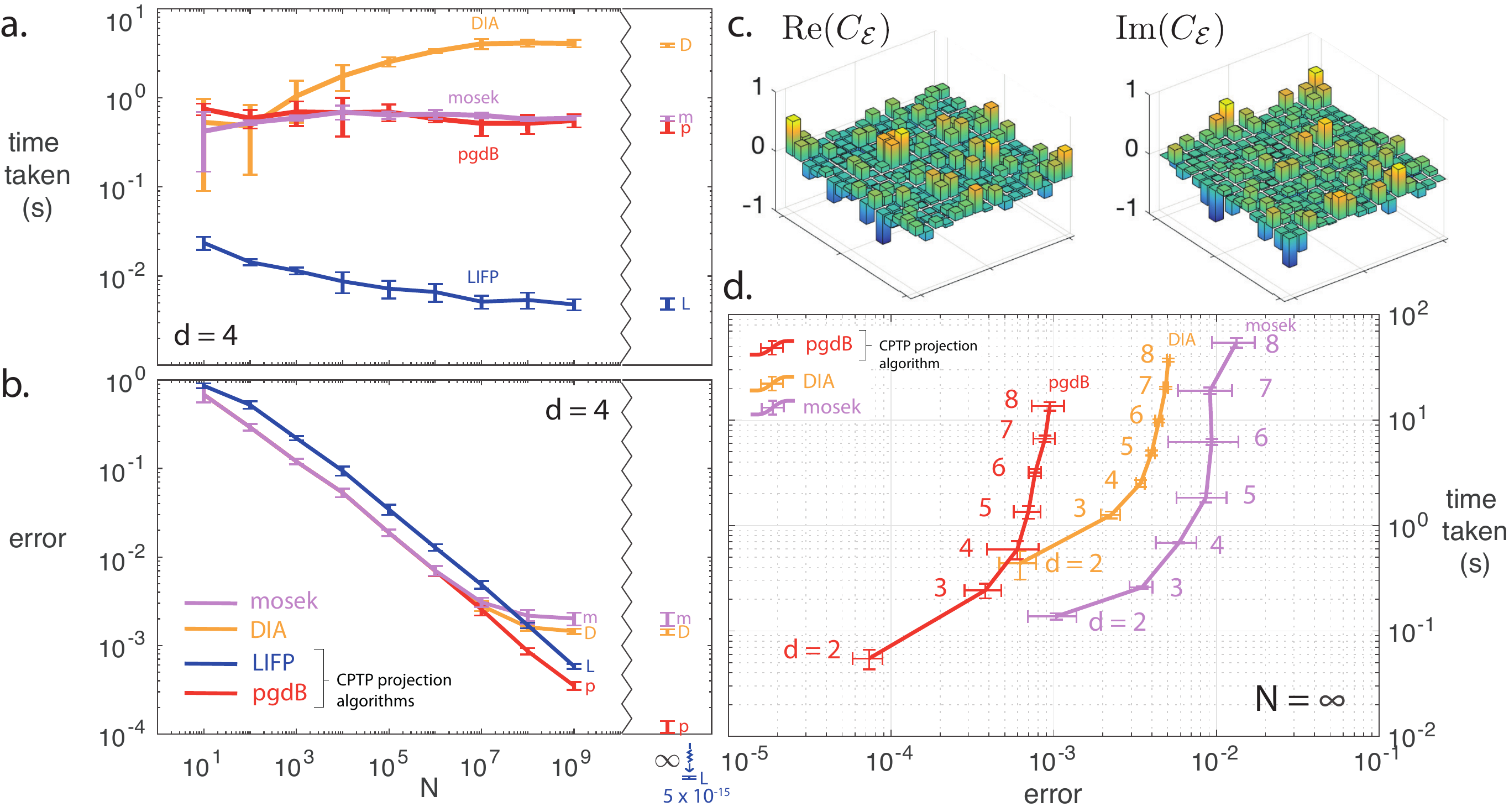}
\caption{\label{combinedfigure}\textbf{a.} Our proposed algorithm pgdB (p, dark red) is significantly faster than the state-of-the-art ML-QPT algorithm DIA (D, light orange) when $N\gtrapprox10^3$. Our other proposed algorithm LIFP (L, dark blue) is faster still, due to its non-iterative nature. 
\textbf{b.} Reconstruction error ($J$ distance) decreases as the number of observations $N$ increases. All of the iterative ML algorithms produce the same quality estimate until $N\gtrapprox10^6$: as statistical noise is made smaller the accuracy of the algorithms deviate, with our proposed algorithms producing much better estimates. We could not simulate arbitrarily high $N$ multinomial noise, but can effectively consider the $N\rightarrow\infty$ case by setting $n_{ij}=p_{ij}$ ($d=4$ case shown). LIFP does not find the ML solution, and therefore sacrifices accuracy in all but the highly idealised case of infinite counts. \textbf{c.} Real and imaginary parts of a typical, purity $>90\%$, $d=4$ Choi matrix returned by our algorithm. \textbf{d.} Accuracy and run time for all iterative ML algorithms as $d$ increases, with negligible statistical noise: pgdB is faster and more accurate. }
\end{figure*}
\section{Numerical benchmarking}
\label{benchmarking}
Unlike other approaches that require an orthonormal operator basis~\cite{SmolinGambettaSmith2012} ours is flexible enough to deal with any informationally complete set of preparations and measurements. To test the performance of our proposed algorithm, we considered the reconstruction of qudit processes, with the $d^2$ preparations $\rho_i=|\psi_i\rangle\langle \psi_i|$ projecting onto the following pure states:
\begin{align}
|\psi_i\rangle = 
\begin{cases}
|j\rangle& j=1\ldots d\\
\frac{|j\rangle+|k\rangle}{\sqrt{2}}& j=1\ldots d,k>j\\
\frac{|j\rangle+\sqrt{-1}|k\rangle}{\sqrt{2}}& j=1\ldots d,k>j
\end{cases}
\end{align}
and a Positive Operator Valued Measure (POVM) formed with the $2d^2$ elements
\begin{align}
E_j = 
\begin{cases}
\rho_i/d^2 & i=1\ldots d^2 \\
\mathbb{I}/d^2-\rho_i/d^2 & i=1\ldots d^2
\end{cases}
\end{align}
which resolve to the identity $\sum_jE_j=\mathbb{I}$. This choice of preparations and measurements is a very practical one, when compared with the use of symmetric informationally complete POVMs (SIC-POVMs)~\cite{RenesBlume-KohoutScott2004} or ancillas~\cite{Leung2003,AltepeterBranningJeffrey2003} which require much greater experimental effort. Appendix~\ref{condition} shows how the choice of measurement and preparation operators affects the conditioning of the optimisation problem. We ignore systematic errors in state preparation and measurement (so-called SPAM errors): gate set tomography~\cite{MerkelGambettaSmolin2013,Blume-KohoutGambleNielsen2013,Blume-KohoutGambleNielsen2017} is a recently proposed solution for a self-consistent characterization in the presence of such imperfections.

We generated random, purity $>90\%$ CPTP maps $C_{\mathcal{E}_{\textrm{true}}}$ in $d=2,\ldots,8$, simulating multinomial random data with sample size $N_i=N$ ranging from $10$ to $10^9$ (see Appendix~\ref{fullrankandquasipure}). Throughout our simulations we normalise the data such that $\sum_j n_{ij}=1$. Although this of little consequence for Maximum Likelihood (ML) and other optimisation approaches (the cost function is merely scaled), it is an important step for Linear Inversion and LIFP, since the latter approach is based on treating the data as probabilities -- hence the need to normalise them into frequencies rather than raw counts.

We also simulated the infinite data case by setting the frequencies equal to the theoretical probabilities $N\rightarrow \infty \Rightarrow n_{ij}=p_{ij}$.  Although not a realistic situation this is a useful device to compare various iterative QPT algorithms as the limit of many observations is approached, where the reconstruction error is dominated by numerical errors rather than noise in the data. We then ran our algorithm (as well as benchmark algorithms) to find $C_{\textrm{est}}$. We recorded running times and accuracies for each. The figure of merit for accuracy was taken to be the $J$ distance, defined as:
\begin{align}
J(\mathcal{E}_{\textrm{est}},\mathcal{E}_{\textrm{true}})= || \mathcal{C}_{\mathcal{E}_{\textrm{est}}}-\mathcal{C}_{\mathcal{E}_{\textrm{true}}}||_\textrm{tr} / 2d.
\end{align}
Here $||\cdot||_\textrm{tr}$ is the trace norm, or sum of singular values, and $J(\mathcal{E}_{\textrm{est}},\mathcal{E}_{\textrm{true}})\in[0,1]$ is related to the average probability of distinguishing the two processes~\cite{GilchristLangfordNielsen2005}. Simulations were run using {\sc matlab} R2017a and {\sc cvx} v2.1 on a Intel Core i7-4790 3.6 GHz with 8 MB L3 cache. 
\subsection{Benchmark 1 : Diluted iterations}
The state-of-the-art algorithm for ML-QPT is the technique of diluted iterations ({\sc dia}), devised by Fiurasek and Hradil in 2001~\cite{Fiur_ek_2001}, who adapted an algorithm for quantum state tomography known as diluted $R\rho R$~\cite{RehacekHradilKnill2007}. The main idea is to exploit an extremal equation obeyed by the ML-CPTP map:
\begin{align}
C_\mathcal{E}=\Lambda^{-1}(C_\mathcal{E})\nabla f (C_\mathcal{E})C_\mathcal{E}\nabla f (C_\mathcal{E})\Lambda^{-1}(C_\mathcal{E}),
\end{align}
where $\Lambda(C_\mathcal{E})=(\textrm{Tr}_{\mathcal{H}_{\textrm{out}}}(C_\mathcal{E}))^{1/2}\otimes\mathbb{I}$ is derived by incorporating the TP constraint through a Lagrange multiplier. This equation is treated as an iterative update rule. The algorithm has been been applied to real data, mostly with Fock-space truncated optical (i.e. continuous variable) systems, according to the formulation by Anis and Lvovsky~\cite{AnisLvovsky2012}. Cooper et al. reported that the algorithm ran for 6.5 hours (with a machine precision stopping rule) to reconstruct a $d=6$ conditional state engineering process~\cite{CooperSladeKarpinski2015}. Fedorov et al. performed QPT of a beamsplitter, revealing the Hong-Ou-Mandel effect~\cite{FedorovFedorovKurochkin2015}. The algorithm typically requires \emph{diluting} with a parameter $0 < \epsilon \leq 1 $ to prevent $f$ from oscillating:
\begin{align}
R^k &= \epsilon\nabla f(C_{\mathcal{E}}^k)+(1-\epsilon)\mathbb{I},\nonumber\\
\Lambda^k &= (\textrm{Tr}_{\mathcal{H}_{\textrm{out}}}(R^kC_{\mathcal{E}}^kR^k))^{1/2}\otimes\mathbb{I},\nonumber\\
C_\mathcal{E}^{k+1}&=(\Lambda^k)^{-1}R^kC_{\mathcal{E}}^kR^k(\Lambda^k)^{-1}.
\end{align}
In our application of {\sc dia}, we used a crude optimisation of $\epsilon$ at each iteration to prevent overshoots. Pseudocode is shown below:

\begin{algorithm}[H]
\caption{{\sc dia}}\label{DIA}
\begin{algorithmic}[1]
\State{$k= 0$}
\State{Initial estimate: $C_0 = \mathbb{I}_{d^2\times d^2}/d$}
\While{$f(C_{\mathcal{E}}^k)-f(C_\mathcal{E}^{k+1})>1\times10^{-10}$ }
\State{$\epsilon = 1$}
\State{$R^k = \epsilon\nabla f(C_{\mathcal{E}}^k)+(1-\epsilon)\mathbb{I}$\;}
\State{$\Lambda^k= (\textrm{Tr}_{\mathcal{H}_{\textrm{out}}}(R^kC_{\mathcal{E}}^kR^k))^{1/2}\otimes\mathbb{I}$}
\While{$f((\Lambda^k)^{-1}R^kC_{\mathcal{E}}^kR^k(\Lambda^k)^{-1})>f(C_{\mathcal{E}})$}
\State{$\epsilon = \epsilon/2$};
\State{$R^k = \epsilon\nabla f(C_{\mathcal{E}}^k)+(1-\epsilon)\mathbb{I}$\;}
\State{$\Lambda^k= (\textrm{Tr}_{\mathcal{H}_{\textrm{out}}}(R^kC_{\mathcal{E}}^kR^k))^{1/2}\otimes\mathbb{I}$}
\EndWhile
\State{$C^{k+1}=(\Lambda^k)^{-1}R^kC_{\mathcal{E}}^kR^k(\Lambda^k)^{-1}$}
\State{$k = k+1$}
\EndWhile
\State{\textbf{Return} $C_{\textrm{est}}= C^{k+1}$}
\end{algorithmic}
\end{algorithm}

\subsection{Benchmark 2: {\sc mosek}}
The problem~\eqref{optiprob} is straightforward to enter into the {\sc cvx} modelling environment~\cite{GrantBoyd2014}, which can then call one of a number of general-purpose solvers.  {\sc cvx} solves problems featuring logarithmic cost functions with a successive approximation method. We found {\sc SeDuMi} failed to find a solution and that {\sc sdpt3} succeeded but was slow. We therefore settled on the commercial solver {\sc mosek}: it usually produced good results but sometimes failed in higher dimensions. We estimate the failure probability to be as high as 15\%, see Appendix~\ref{app:mosek}.

\section{Results and conclusions} 
\label{conclusions}
Our results are summarised in Fig.~\ref{combinedfigure}: The CPTP-projection-based algorithms are not only faster than existing approaches, but significantly more accurate. This is especially the case as $d$ increases but is true even for $d=4$ (corresponding, for example, to a 2-qubit gate such as the controlled-NOT). 

We confirmed that due to the iterative nature of pgdB, higher accuracies can be achieved by adjusting the stopping criterion and other metaparameters, at the expense of a longer algorithm runtime.  LIFP is a quick and effective method suited to situations with low noise and low $d$, and sacrificing a small amount of accuracy, typically less than an order of magnitude, with respect to the iterative methods. Appendix~\ref{fullrankandquasipure} includes results corresponding to reconstruction of full rank processes, which show qualitatively similar trends. 

Our results indicate that CPTP projection is a valuable and versatile tool for QPT, holding great promise for applications where time and accuracy are both important. {\sc matlab} code is available at \url{https://github.com/geoknee/CPTPprojection}.

\begin{acknowledgments}
 G.C.K was supported by the Royal Commission for the Exhibition of 1851 and thanks Yanbao Zhang for discussions. E.M.G. thanks the Royal Society of Edinburgh and the Scottish Government for support. We thank Yen-Huan Li for useful feedback on this document. 
 \end{acknowledgments}

\appendix
\section{Vectorizing the problem}
\label{app:vector}
Our forward model may be vectorised as
$
p := \texttt{vec}[p_{ij}]=A\texttt{vec}[C_\mathcal{E}]
$
upon defining
\begin{align}
A = 
\left(
\begin{array}{c}
\texttt{vec}[\rho_1^*\otimes E_1^\dagger]^\dagger\\
\vdots\\
\texttt{vec}[\rho_1^*\otimes E_N^\dagger]^\dagger\\
\vdots\\
\texttt{vec}[\rho_N^*\otimes E_N^\dagger]^\dagger
\end{array}
\right)=
\left(
\begin{array}{c}
\texttt{vec}[\rho_1\otimes E_1^T]^T\\
\vdots\\
\texttt{vec}[\rho_1\otimes E_N^T]^T\\
\vdots\\
\texttt{vec}[\rho_N\otimes E_N^T]^T
\end{array}
\right).
\label{Adef}
\end{align}
Now $f(C_\mathcal{E})=-n^T \ln(p)$ and $\nabla f(C_\mathcal{E}) = - A^\dagger\eta$ with $\eta_{ij} = n_{ij}/p_{ij}$ and $\eta=\texttt{vec}[\eta_{ij}]$ and so on. These facts follow from elementary matrix calculus along with the identity $\text{Tr}(A^\dagger B) \equiv \texttt{vec}[A]^\dagger\texttt{vec}[B]$. Shang et al. present a method to speedup calculation of the gradient when $A$ has a tensor product structure~\cite{ShangZhangNg2017}: unfortunately this does not help in our case because we have only a single tensor product $\rho_i^T\otimes E_j$ and multiple tensor products are needed to show an advantage. In a slight abuse of notation, above we have denoted the vectorization of a matrix $M$ by $\texttt{vec}[M_{ij}]$. where $M_{ij}$ are the matrix elements of $M$.
\section{Trace preservation constraint}
\label{app:tp}
Counter to a common misconception, the celebrated `Choi-Jamiolokowski isomorphism'~\cite{JiangLuoFu2013} does not imply a 1:1 correspondence between CPTP maps and density operators $C_{\mathcal{E}}/\textrm{Tr}(C_{\mathcal{E}})$ in a higher dimensional space. As discussed in the main paper, 
some such density operators do not correspond to CPTP maps~\cite{JiangLuoFu2013}. See for example Refs~\cite{Sacchi2001,FiurasekHradil2001}. As an example of why a stronger condition is important, consider the Choi matrix $C_{\Box}=\textrm{{dia}g}(0.1,0.1,0.1,1.7)$. This is proportional to a perfectly valid, classically correlated 2-qubit state. It is manifestly positive, and has trace equal to 2. Note however, that $\textrm{Tr}_2(C_{\Box})=\textrm{{dia}g}(0.2,1.8)\neq \mathbb{I}.$ This means that it does not represent a trace preserving map. Note that the given example actually increases the trace of some states and decreases the trace of others so is neither in TP nor in TNI (discussed below).%

\section{Projection onto the set of trace non-increasing processes}
\label{app:tni}
It will be possible to relax our algorithmic projections in such a way as to search in larger spaces, i.e. supersets of TP. This may be important for certain applications: in fact, whenever there is loss or some other non deterministic process such as a measurement, CPTP maps (or `channels') are replaced by CPTNI maps (or  `operations') which are \emph{trace non-increasing}. The trace of the output density matrix corresponds to the probability of success of the map. CP maps admit a Kraus representation~\cite{Choi1975}:
\begin{align}
\mathcal{E}(\rho)=\sum_i K_i\rho K_i^\dagger.
\end{align}
The difference in trace between an input and an output state is given by
\begin{align}
\textrm{Tr}(\mathcal{E}(\rho))-\textrm{Tr}(\rho)&=\textrm{Tr}(\sum_i K_i\rho K_i^\dagger)-\textrm{Tr}(\rho)\nonumber\\
&=\textrm{Tr}([\mathcal{Y} -\mathbb{I}]\rho),
\end{align}
where we used the cyclic property of the trace and defined $\mathcal{Y}=\sum_i K_i^\dagger K_i$. Since $\rho$ is an arbitrary positive operator, it is clear that trace preservation is equivalent to
\begin{align}
\mathcal{Y}=\mathbb{I}\qquad\textrm{(TP)}.
\end{align}
 In fact, it is always possible to diagonalise $\mathcal{Y}$ with a similarity transformation $\mathcal{V}$. Let $\lambda_i$ be the eigenvalues of $\mathcal{Y}$~\cite{BongioanniSansoniSciarrino2010}. Then
 \begin{align}
\textrm{Tr}(\mathcal{E}(\rho))-\textrm{Tr}(\rho) = \textrm{Tr}(\textrm{diag}[\lambda_1-1,\lambda_2-1,\ldots]\mathcal{V}^\dagger\rho\mathcal{V}).
\label{basisindependent}
\end{align}
So a general CP map will change the trace of an input state depending on the eigenvalues of $\mathcal{Y}$, and on the projection of the state onto the eigenvectors of $\mathcal{Y}$.
We can consider the case where the process has a uniform and known success probability $p$:
\begin{align}
\mathcal{Y}=p\mathbb{I}\qquad(\textrm{US}_p).
\end{align}
The projection onto this set is easy: simply alter the TP projection given in Eq.~\eqref{tp_projection} by taking ${\texttt{vec}[\mathbb{I}]}\rightarrow p{\texttt{vec}[\mathbb{I}]}$. 
 If we demand that the trace is non-increasing for an arbitrary input state, we have
\begin{align}
\mathcal{Y}\preceq\mathbb{I}\qquad\textrm{(TNI)}.
\label{TNI}
\end{align}

A commonly used approach to reconstruct TNI operations involves introducing of a fictitious `shelving' state (thereby extending the space $d\rightarrow d+1$~\cite{AnisLvovsky2012}). One then performs a TP reconstruction in the larger space before projecting out the fictitious state. 
We present an alternative here that avoids (the possibly very expensive downside of) having to increase the size of the Hilbert space. By the definition of the Choi matrix, 
\begin{align}
\textrm{Tr}_{\mathcal{H}_{\textrm{out}}}(C_{\mathcal{E}})=\mathcal{Y}.
\end{align}
It is straightforward to project onto TNI (which is a convex superset of TP). The correlations $\chi$ in the Choi matrix are implicitly defined through
\begin{align}
C_{\mathcal{E}} = \textrm{Tr}_{\mathcal{H}_{\textrm{out}}}(C_{\mathcal{E}})\otimes\mathbb{I}+\mathbb{I}\otimes\textrm{Tr}_{\mathcal{H}_{\textrm{in}}}(C_{\mathcal{E}})+\chi.
\end{align}
We can achieve the TNI projection via 

\begin{align}
\mathcal{TNI}(C_{\mathcal{E}}) & =
C_{\mathcal{E}}+\frac{1}{d}[\mathcal{G}(\textrm{Tr}_{\mathcal{H}_{\textrm{out}}}(C_{\mathcal{E}}))-\textrm{Tr}_{\mathcal{H}_{\textrm{out}}}(C_{\mathcal{E}})]\otimes\mathbb{I}.
\end{align}
where we introduced $\mathcal{G}$, defined and computed via
\begin{align}
\mathcal{G}(X) &= \arg\min_{\textrm{Tr}_{\mathcal{H}_{\textrm{out}}}(B)\preceq\mathbb{I}}||\textrm{Tr}_{\mathcal{H}_{\textrm{out}}}(B)\otimes\mathbb{I}-X\otimes\mathbb{I}||\\
&= \arg\min_{\textrm{Tr}_{\mathcal{H}_{\textrm{out}}}(B)\preceq\mathbb{I}}||\textrm{Tr}_{\mathcal{H}_{\textrm{out}}}(B)-X||\cdot ||\mathbb{I}||\\
&= \arg\min_{\textrm{Tr}_{\mathcal{H}_{\textrm{out}}}(B)\preceq\mathbb{I}}||\textrm{Tr}_{\mathcal{H}_{\textrm{out}}}(B)-X||\\
&= \arg\min_{E\preceq\mathbb{I}}||E-X||\\
&= V\min(D,1)\, V^\dagger
\end{align}
with $X=VDV^\dagger$~\cite{LewisMalick2008}. To see that we have the orthogonal projection, consider that for any $B\in$ TNI we may write $B =\textrm{Tr}_{\mathcal{H}_{\textrm{out}}}(B)\otimes\mathbb{I}+\mathbb{I}\otimes\textrm{Tr}_{\mathcal{H}_{\textrm{in}}}(B)+\chi_B $. Now
\begin{widetext}
\begin{align}
||B-\mathcal{E}|| &= ||  \textrm{Tr}_{\mathcal{H}_{\textrm{out}}}(B-C_{\mathcal{E}})\otimes\mathbb{I}+\mathbb{I}\otimes\textrm{Tr}_{\mathcal{H}_{\textrm{in}}}(B-C_{\mathcal{E}})+\chi_B-\chi||\\
&\leq ||  \textrm{Tr}_{\mathcal{H}_{\textrm{out}}}(B-C_{\mathcal{E}})\otimes\mathbb{I}||+||\mathbb{I}\otimes\textrm{Tr}_{\mathcal{H}_{\textrm{in}}}(B-C_{\mathcal{E}})||+||\chi_B-\chi||\\
&\leq ||  \textrm{Tr}_{\mathcal{H}_{\textrm{out}}}(B-C_{\mathcal{E}})\otimes\mathbb{I}||\leq \min_{\textrm{Tr}_{\mathcal{H}_{\textrm{out}}}(B)\preceq\mathbb{I}} ||  \textrm{Tr}_{\mathcal{H}_{\textrm{out}}}(B)\otimes\mathbb{I}-\textrm{Tr}_{\mathcal{H}_{\textrm{out}}}(C_{\mathcal{E}})\otimes\mathbb{I}||\\
&= \min_{\textrm{Tr}_{\mathcal{H}_{\textrm{out}}}(B)\preceq\mathbb{I}} ||  \textrm{Tr}_{\mathcal{H}_{\textrm{out}}}(B)-\textrm{Tr}_{\mathcal{H}_{\textrm{out}}}(C_{\mathcal{E}})||\cdot ||\mathbb{I}||=\min_{\textrm{Tr}_{\mathcal{H}_{\textrm{out}}}(B)\preceq\mathbb{I}} \frac{1}{d}||  \textrm{Tr}_{\mathcal{H}_{\textrm{out}}}(B)-\textrm{Tr}_{\mathcal{H}_{\textrm{out}}}(C_{\mathcal{E}})||\\
&= ||\mathcal{TNI}(C_{\mathcal{E}})-C_{\mathcal{E}}||,
\end{align}
\end{widetext}
and therefore $\mathcal{TNI}(C_{\mathcal{E}})$ is the closet element in TNI to $C_{\mathcal{E}}$. We verified this numerically using {\sc cvx}.
Using the techniques shown in the main text, this result enables tomography of quantum operations belonging to CPTNI without expanding the reconstruction space.
\section{Stalling}
\label{stalling}
\begin{figure*}
\includegraphics[width=\textwidth]{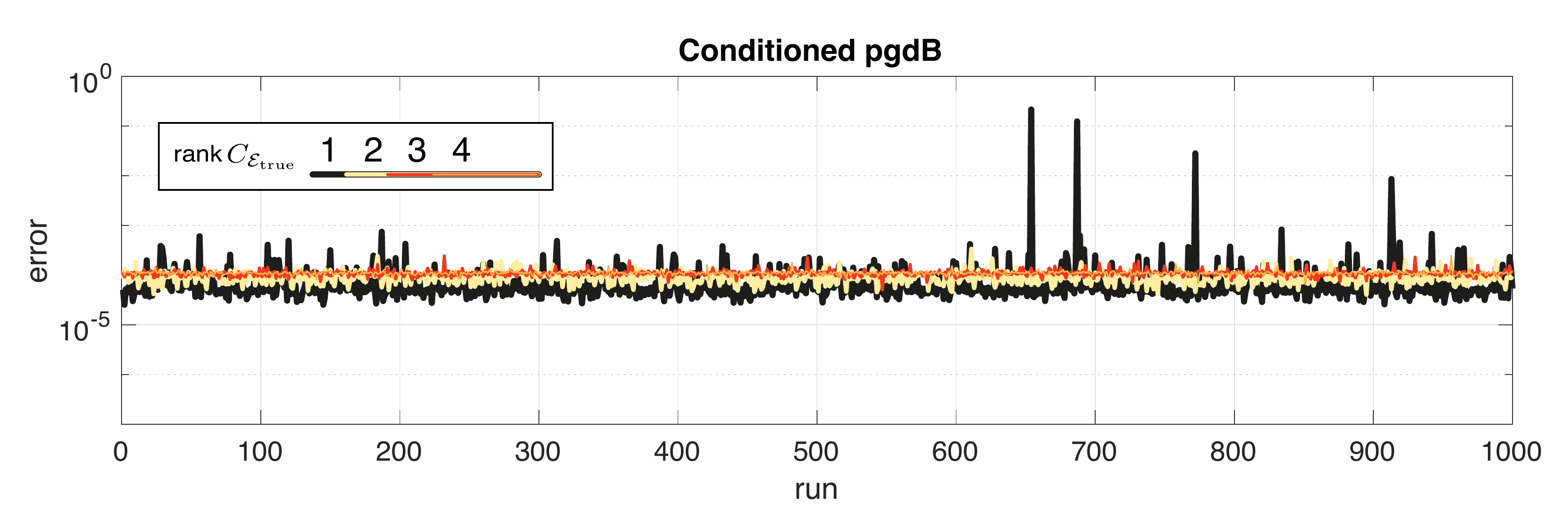}
\caption{\label{pgdb_robust}Our conditioning step means the pgdB algorithm is robust to reconstruction of rank deficient processes, even with a multinomial likelihood function. The algorithm never stalls but in rare cases the accuracy is adversely affected. Here $d=2$ and $\epsilon=1\times10^{-16}$. }
\end{figure*}
As mentioned in the main text, it is possible for pgdB to stall if a situation with $p_{ij}=0$ is encountered, and in this section we discuss a simple solution allowing the algorithm to continue. Importantly, our proposed solution alerts the user to when the stalling fix has been applied, turning such a rare occurrence into a heralded event. It is important to note that other potential solutions also exist, but a comprehensive study of their merits and drawbacks (including bounding any systemic inaccuracies that arise) are beyond the scope of the present work.

One straightforward solution to keeping the algorithm running when a $p_{ij}=0$ situation arises is to modify the cost function and gradient at each iteration with the heralded conditioning step:
\begin{align}
p_{ij}=\max(p_{ij},\epsilon)
\label{conditioning}
\end{align}
for $\epsilon$ some small parameter. It is clearly possible that the algorithm never encounters $p_{ij}<\epsilon$; in that case our conditional modification does not occur and we proceed toward the ML solution. In the case that the modification is triggered, the algorithm is capable of raising a flag, warning the user to interpret the resultant solution accordingly -- it would also be possible for the flag to trigger a restart of the algorithm with a random initial guess. In theory, the ML solution cannot have any $p_{ij}=0$ (since that is a zero of the likelihood, which is positive semidefinite). The exception is when $p_{ij}=n_{ij}=0$, trivial cases excluded by taking the convention $0^0=1$. Therefore 
\begin{align}
p_{ij}=
\begin{cases}
p_{ij},\qquad &p_{ij}>0\\
\epsilon,\qquad &p_{ij}=0
\end{cases}
\end{align}
is sufficient to keep the algorithm going without altering the turning points of the objective function. In practice, machines with finite precision cannot calculate the logarithm of very small numbers, necessitating the stronger conditioning step given in Eq.~\eqref{conditioning}. The implication is that, while the ML solution cannot be at $p_{ij}=0$, it can have $p_{ij}<\epsilon$. Therefore with low probability the conditioning step can affect the accuracy of the reconstructed solution (but this possibility is heralded). The probability of this occurring is related to the Kraus rank of the process -- see Fig.~\ref{pgdb_robust}, which shows that the conditioning step prevents the stalling problem with no impact on the reconstruction error in most cases. 
\section{Linear Inversion with a final projection onto CPTP}
\label{app:LIFP}
\begin{figure*}
\includegraphics[width=0.8\textwidth]{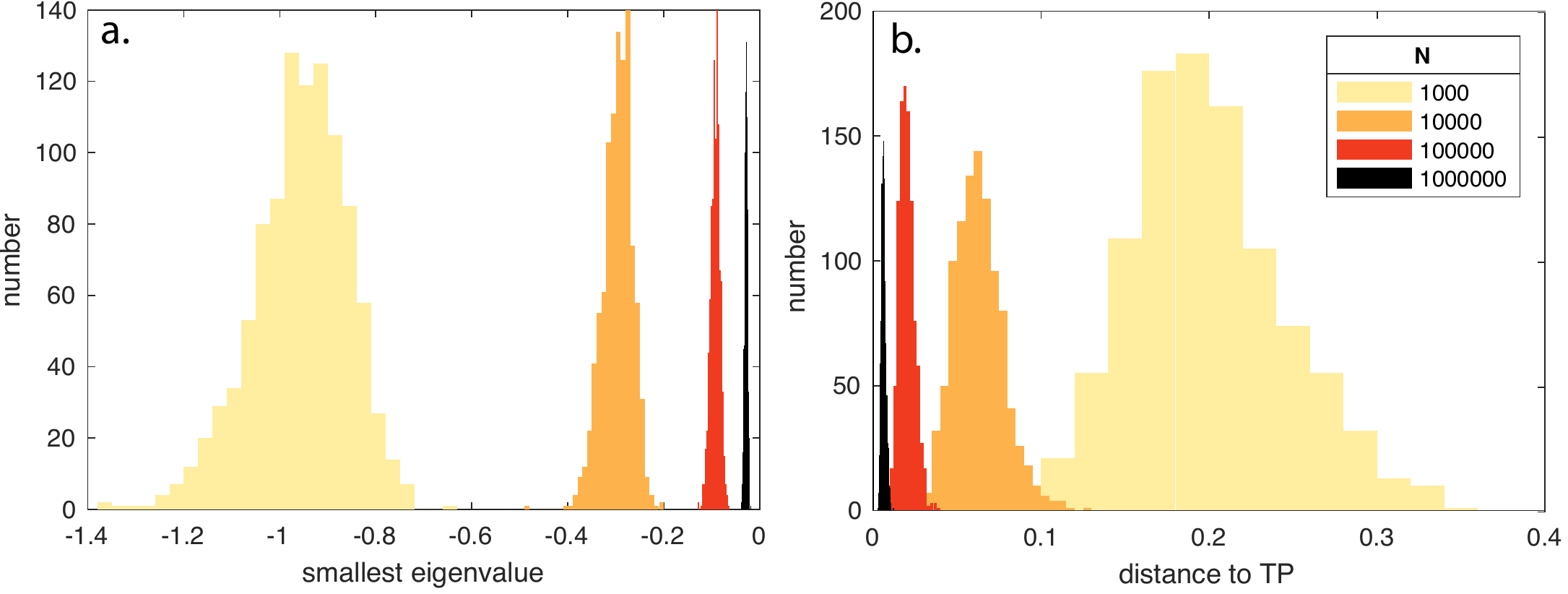}
\caption{Distribution of \textbf{a.} the smallest eigenvalue of $C^{LI}_\mathcal{E}$ and \textbf{b.} distance $||\textrm{Tr}_2(C_{\mathcal{E}}^{LI})-\mathbb{I}||_2$  to the TP set under various degrees of multinomial shot noise in $d=4$ with an ensemble of full rank CPTP maps. As the number of clicks $N$ increases, the typical magnitude of the smallest eigenvalue decreases, as does the distance to TP. Unphysical estimates (e.g. where the smallest eigenvalue is negative) occur in almost all cases (1000 sampled for each value of $N$). \label{smallestEV}}
\end{figure*}
The most straightforward approach to inverting the forward model is arguably the approach of linear inversion. Here, the estimate is 
\begin{align}
C^{LI}_\mathcal{E}:= {\tt vec}^{-1} [A^{+}n],
\end{align}
where $A^+=A^\dagger(AA^\dagger)^{-1}$ is the pseudo-inverse of $A$~\cite{Teo2015}, and $A$ is defined in Eq.~\eqref{Adef}. This estimator minimises the distance in Frobenius norm between the data and the forward model -- a familiar notion of `least squares' fitting. This notion of discrepancy is distinct from the negative likelihood used in the main paper. Note however, that in the special case where the likelihood is a Gaussian function, or indeed when there is no noise at all, the two notions of discrepancy coincide~\cite{SmolinGambettaSmith2012}. Linear Inversion is a non-iterative approach so is likely to be fast in low dimensions, although the number of floating point operations will be $O(d^{12})$ and will therefore not scale as well as pgdB (which is $O(d^8)$). Furthermore without the explicit expression of the CP and TP constraints, there is no guarantee of a physical estimate. 

We implemented this estimator using the {\sc matlab} backslash operator $\backslash$, which avoids explicit calculation of the pseudo-inverse and instead uses a method of Gaussian elimination. We projected the estimate onto the space of Hermitian matrices $C^{LI}_\mathcal{E}\rightarrow(C^{LI}_\mathcal{E}+(C^{LI}_\mathcal{E})^\dagger)/2 $ in order to investigate their subsequently real eigenspectrum.  We found that estimates were unphysical with overwhelming probability -- see Fig.~\ref{smallestEV}. The severity of the problem increases with shot noise. It is therefore not possible to use the standard analytical tools of infidelity or $J$ distance to meaningfully compare the accuracy of linear inversion with ML. It is worth noting that the ML approach provides estimates with smallest eigenvalue no less than some negative quantity which can be set arbitrarily small. 
\begin{figure}
\centering
\includegraphics[width=\columnwidth]{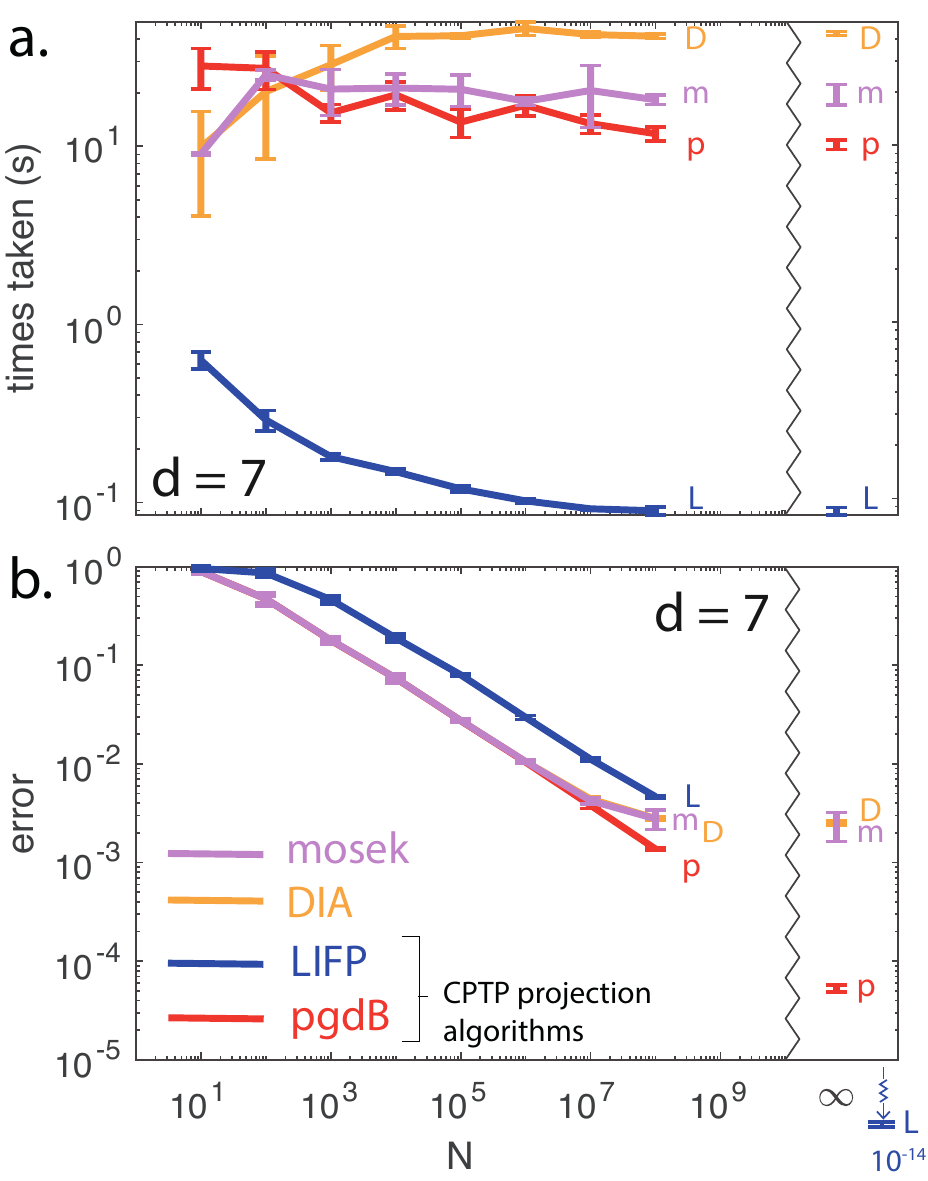}
\caption{Performance (in terms of \textbf{a.} time and \textbf{b.} accuracy) of Linear Inversion with a Final Projection (LIFP) compared to the ML techniques, for $d=7$ quasi-pure processes. When there is no noise whatsoever $N\rightarrow\infty$, the final projection is not necessary and the error is very low (consistent with the product of the machine precision with the condition number of $A$). In all other cases shown, a significant drop in precision may be seen. \label{LIFP}}
\end{figure}

To allow for a comparison with the manifestly physical approach of ML, we upgraded linear inversion to force it to also produce physical estimates --- simply by reapplying the novel, composite CPTP projection derived in the main paper just once. We call the resultant estimator Linear Inversion with a Final Projection:
\begin{align}
C^{LIFP}_\mathcal{E}:= \mathcal{CPTP}( {\tt vec}^{-1} [A^{+}n]),
\end{align}
once more forgoing explicit calculation of $A^+$ in favour of the {\sc matlab} backslash operator $\backslash$. This approach is very much in the same spirit as the `quick and dirty' state tomography approach used by Kaznady and James~\cite{KaznadyJames2009}. We supplement the results in the main paper with an investigation in $d=7$, shown in Fig.~\ref{LIFP}. The results confirm the idea that linear inversion with a final projection is a quick and effective method when the number of trials is taken to infinity, but that the loss in accuracy for finite $N$ is exacerbated in higher dimensions. In fact, the projection is not necessary when there is zero noise. When noise is negligible, linear inversion provides a very high precision estimate due to its non-iterative nature, but our results suggest that ML approaches this situation faster as multinomial noise is decreased. Otherwise, in the physically relevant scenario of finite $N$ a significant loss in precision is clear to see. The running time of Linear Inversion with a Final Projection can be favourable in low dimensions but will become much slower than the other methods as $d$ increases.

\section{Condition number of $A$}
\label{condition}
\begin{figure}[t!]
\includegraphics[width=8cm]{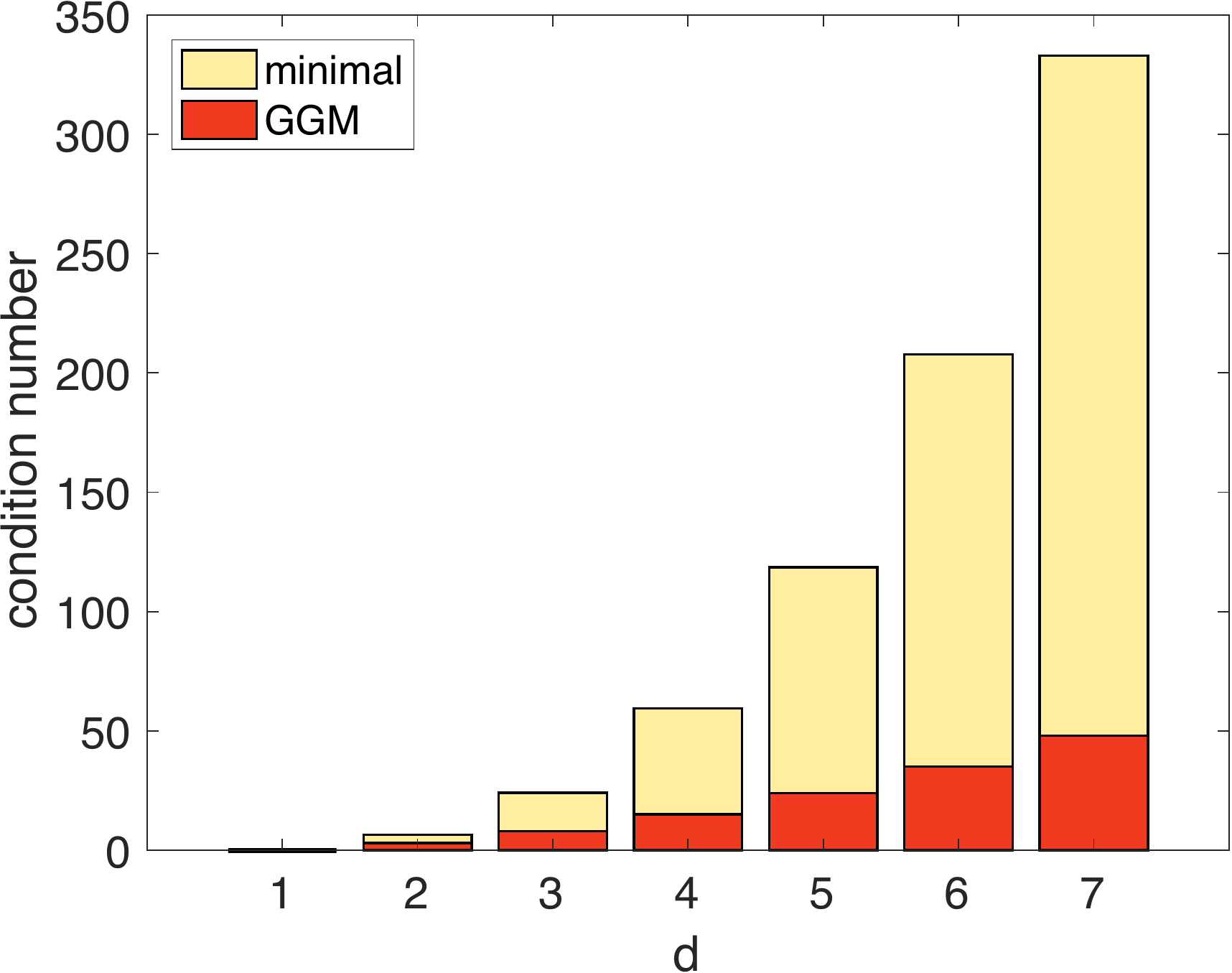}
\caption{\label{cond}Condition number of  $A$ as a function of $d$ for our chosen `minimal' set of preparations and measurements (blue) and for a set constructed from all eigenvectors of the generalised Gell-Mann matrices (GGM, red). }
\end{figure}
Fig~\ref{cond} shows the condition number of $A$ rising with $d$, for the `minimal' choice of preparations and measurements made in the main text. A high condition number means a slower runtime and a less accurate result in general. The condition number might be brought down by considering over-complete preparations and measurements, or SIC-POVMs~\cite{RenesBlume-KohoutScott2004}.  The density of $A$ is typically around 10 percent or less, meaning efficiency savings for sparse operations with {\sc matlab}. 

The computational complexity of the gradient-based algorithms is dominated by the matrix-vector product required to calculate the gradient. The complexity is $\mathcal{O}(d^4n_{\textrm{combs}})$ where $n_{\textrm{combs}}$ is the number of combinations of preparations and measurement outcomes. Our choice above implies $n_{\textrm{combs}}=2d^4$. 
An over-complete set would impact running time: for example, choosing the eigenvectors of the generalised Gell-Mann matrices~\cite{BertlmannKrammer2008} implies $n_{\textrm{combs}}=(d^3-d)^2$, and raises the complexity from $\mathcal{O}(d^{8})$ to $\mathcal{O}(d^{10})$.  

\section{Failure of {\sc cvx}}
\begin{figure}
\includegraphics[width=\columnwidth]{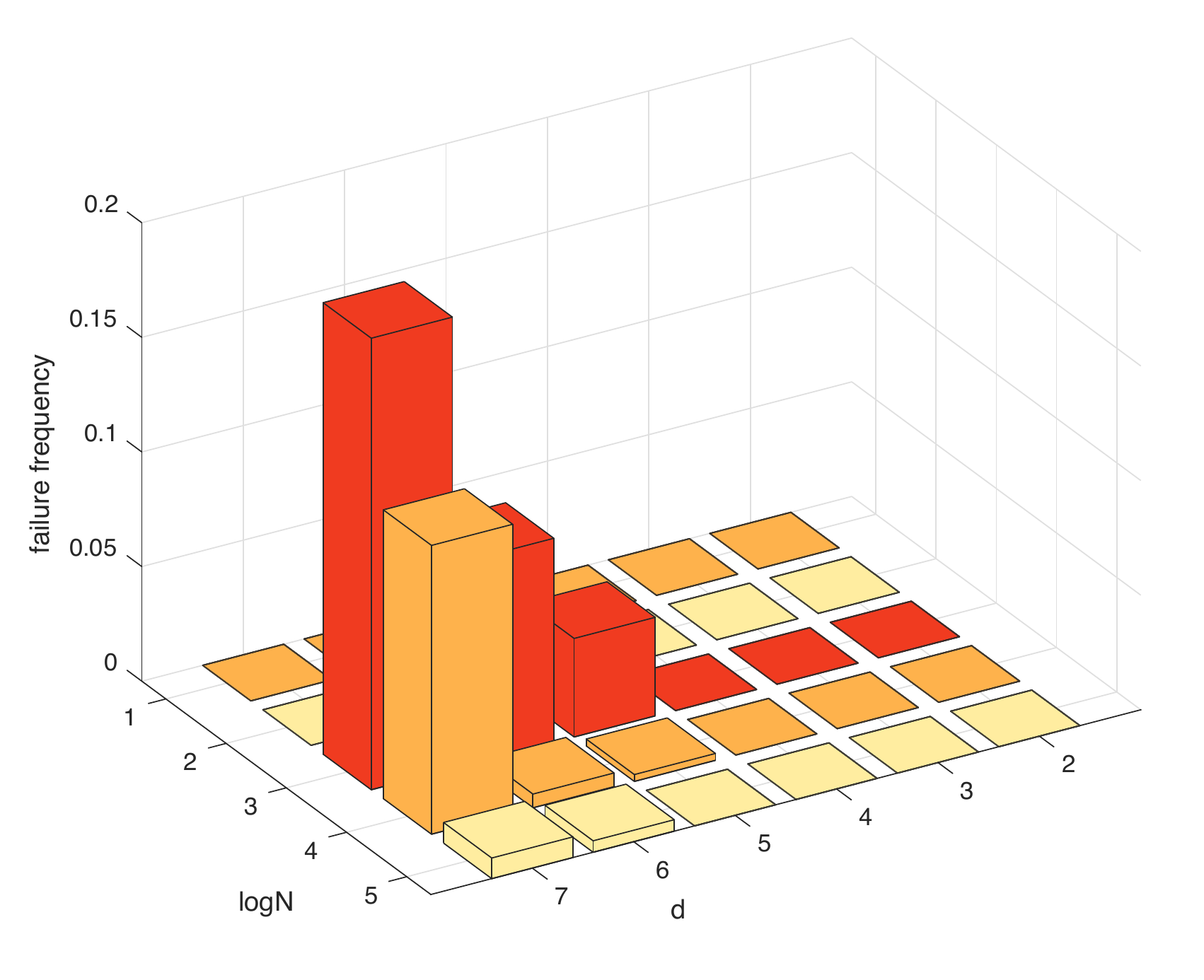}
\caption{Failure frequency of {\sc cvx} and the {\sc mosek} solver, depending on multinomial statistical noise $\log_{10}N$ and Hilbert space dimension $d$. 1000 points were sampled for each combination of $N$ and $d$. \label{failure}  }
\end{figure}
\label{app:mosek}
Depending on dimensionality and statistical noise, we found the experimental method of successive approximation used by {\sc cvx}~\cite{GrantBoyd2014} to be unreliable. This is a known issue and to be expected due to the heuristic nature of the method. Fig.~\ref{failure} shows the failure rate to be as high as 15\%. The data for {\sc mosek} used in Figs.~\ref{combinedfigure}, \ref{LIFP} and \ref{FULLRANKcombinedfigure} are post-selected on the algorithm succeeding.
\clearpage
\begin{widetext}
\section{Generation of full-rank and quasi-pure random CPTP maps}
\label{fullrankandquasipure}
Random CPTP maps may be uniformly randomly generated according to a prescription by Bruzda et al.~\cite{BruzdaCappelliniSommers2009}:
\begin{align}
Y &= (\textrm{Tr}_{\mathcal{H}_{\textrm{out}}}[XX^\dagger])^{1/2},\\
B&=(\mathbb{I}\otimes Y^{-1/2})XX^\dagger(\mathbb{I}\otimes Y^{-1/2}),
\end{align}
where $X$ is a $d^2\times M$ complex random matrix with entries distributed normally. The Kraus rank of the CPTP map is $M$. We generated $d^2$ rank-1 matrices $B_i$ and formed their convex combination $C_{\mathcal{E}_{\textrm{true}}}=\sum_iP_iB_i$ with an exponentially decaying probability distribution $P_i$ with  $\sum_i P_i^2=0.9$. The resulting ensemble is CPTP (by convexity) and has purity $\textrm{Tr}[C_{\mathcal{E}}^2]/d^2$ no less than 90\%. We call this the `quasi-pure' ensemble.

Using the first part of the above procedure we generated an alternative data set corresponding to full rank processes. The results are shown in Fig.~\ref{FULLRANKcombinedfigure}, where the lower purity of this ensemble (compared to the quasi-pure ensemble of Fig.~\ref{combinedfigure}) is apparent from the relative closeness of a typical Choi matrix to $\textrm{diag}(1,1,\ldots 1)/d$.  Our proposed algorithm is still superior to existing methods, although less dramatically so. 

\begin{figure*}
\includegraphics[width=\textwidth]{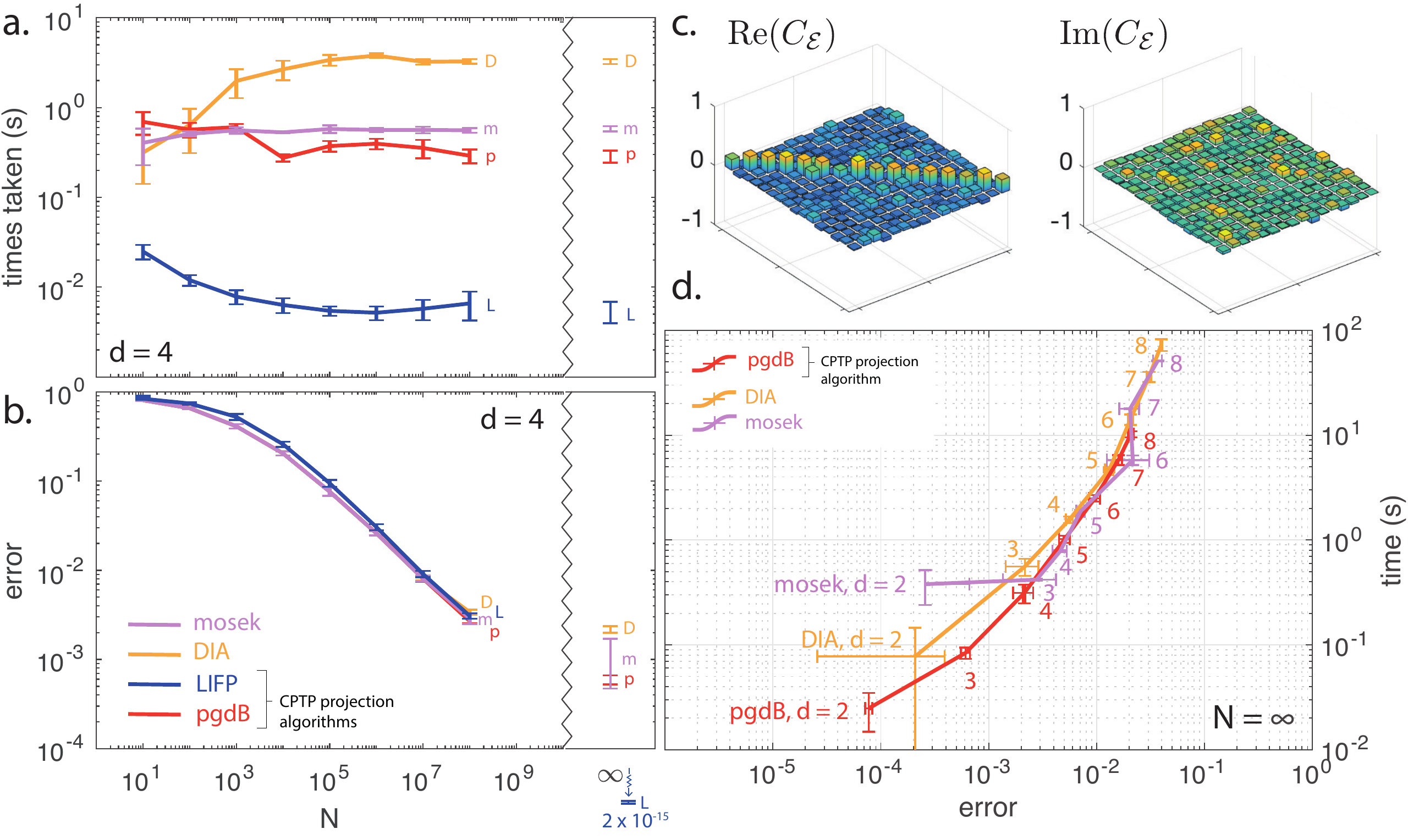}
\caption{\label{FULLRANKcombinedfigure} Reconstruction of full-rank processes \textbf{a.} Our proposed algorithm pgdB (p dark red) is significantly faster than the state-of-the-art ML-QPT algorithm DIA (D, light orange) when $N\gtrapprox10^3$. LIFP (L, dark blue) is faster still, due to its non-iterative nature. 
\textbf{b.} Reconstruction error ($J$ distance) decreases as the number of observations $N$ increases. All algorithms produce approximately the same quality estimate. We could not simulate arbitrarily high $N$ multinomial noise, but can effectively consider the $N\rightarrow\infty$ case by setting $n_{ij}=p_{ij}$ , where pgdB becomes the most accurate ML algorithm ($d=4$ case shown). LIFP does not find the ML solution, and therefore sacrifices accuracy in all but the highly idealised case of infinite counts. \textbf{c.} Real and imaginary parts of a typical, full rank, $d=4$ Choi matrix returned by our algorithm. \textbf{d.} Accuracy and run time for all iterative ML algorithms as $d$ increases, with negligible statistical noise: pgdB is faster and more accurate.  }
\end{figure*}
\end{widetext}
\clearpage
\bibliography{/Users/georgeknee/Documents/paper_library/gck_full_bibliography}

\begin{thebibliography}{50}%
\makeatletter
\providecommand \@ifxundefined [1]{%
 \@ifx{#1\undefined}
}%
\providecommand \@ifnum [1]{%
 \ifnum #1\expandafter \@firstoftwo
 \else \expandafter \@secondoftwo
 \fi
}%
\providecommand \@ifx [1]{%
 \ifx #1\expandafter \@firstoftwo
 \else \expandafter \@secondoftwo
 \fi
}%
\providecommand \natexlab [1]{#1}%
\providecommand \enquote  [1]{``#1''}%
\providecommand \bibnamefont  [1]{#1}%
\providecommand \bibfnamefont [1]{#1}%
\providecommand \citenamefont [1]{#1}%
\providecommand \href@noop [0]{\@secondoftwo}%
\providecommand \href [0]{\begingroup \@sanitize@url \@href}%
\providecommand \@href[1]{\@@startlink{#1}\@@href}%
\providecommand \@@href[1]{\endgroup#1\@@endlink}%
\providecommand \@sanitize@url [0]{\catcode `\\12\catcode `\$12\catcode
  `\&12\catcode `\#12\catcode `\^12\catcode `\_12\catcode `\%12\relax}%
\providecommand \@@startlink[1]{}%
\providecommand \@@endlink[0]{}%
\providecommand \url  [0]{\begingroup\@sanitize@url \@url }%
\providecommand \@url [1]{\endgroup\@href {#1}{\urlprefix }}%
\providecommand \urlprefix  [0]{URL }%
\providecommand \Eprint [0]{\href }%
\providecommand \doibase [0]{http://dx.doi.org/}%
\providecommand \selectlanguage [0]{\@gobble}%
\providecommand \bibinfo  [0]{\@secondoftwo}%
\providecommand \bibfield  [0]{\@secondoftwo}%
\providecommand \translation [1]{[#1]}%
\providecommand \BibitemOpen [0]{}%
\providecommand \bibitemStop [0]{}%
\providecommand \bibitemNoStop [0]{.\EOS\space}%
\providecommand \EOS [0]{\spacefactor3000\relax}%
\providecommand \BibitemShut  [1]{\csname bibitem#1\endcsname}%
\let\auto@bib@innerbib\@empty
\bibitem [{\citenamefont {Aharonov}\ and\ \citenamefont
  {Ben-Or}(1997)}]{AharonovBen-Or1997}%
  \BibitemOpen
  \bibfield  {author} {\bibinfo {author} {\bibfnamefont {D.}~\bibnamefont
  {Aharonov}}\ and\ \bibinfo {author} {\bibfnamefont {M.}~\bibnamefont
  {Ben-Or}},\ }\bibfield  {title} {\enquote {\bibinfo {title} {Fault-tolerant
  quantum computation with constant error},}\ }in\ \href {\doibase
  10.1145/258533.258579} {\emph {\bibinfo {booktitle} {Proceedings of the
  Twenty-ninth Annual ACM Symposium on Theory of Computing}}},\ \bibinfo
  {series and number} {STOC '97}\ (\bibinfo  {publisher} {ACM},\ \bibinfo
  {address} {New York, NY, USA},\ \bibinfo {year} {1997})\ pp.\ \bibinfo
  {pages} {176--188}\BibitemShut {NoStop}%
\bibitem [{\citenamefont {O'Brien}\ \emph {et~al.}(2004)\citenamefont
  {O'Brien}, \citenamefont {Pryde}, \citenamefont {Gilchrist}, \citenamefont
  {James}, \citenamefont {Langford}, \citenamefont {Ralph},\ and\ \citenamefont
  {White}}]{OBrienPrydeGilchrist2004}%
  \BibitemOpen
  \bibfield  {author} {\bibinfo {author} {\bibfnamefont {J.~L.}\ \bibnamefont
  {O'Brien}}, \bibinfo {author} {\bibfnamefont {G.~J.}\ \bibnamefont {Pryde}},
  \bibinfo {author} {\bibfnamefont {A.}~\bibnamefont {Gilchrist}}, \bibinfo
  {author} {\bibfnamefont {D.~F.~V.}\ \bibnamefont {James}}, \bibinfo {author}
  {\bibfnamefont {N.~K.}\ \bibnamefont {Langford}}, \bibinfo {author}
  {\bibfnamefont {T.~C.}\ \bibnamefont {Ralph}}, \ and\ \bibinfo {author}
  {\bibfnamefont {A.~G.}\ \bibnamefont {White}},\ }\bibfield  {title} {\enquote
  {\bibinfo {title} {Quantum process tomography of a controlled-not gate},}\
  }\href {\doibase 10.1103/PhysRevLett.93.080502} {\bibfield  {journal}
  {\bibinfo  {journal} {Phys. Rev. Lett.}\ }\textbf {\bibinfo {volume} {93}},\
  \bibinfo {pages} {080502} (\bibinfo {year} {2004})}\BibitemShut {NoStop}%
\bibitem [{\citenamefont {Pach{\'o}n}\ \emph {et~al.}(2015)\citenamefont
  {Pach{\'o}n}, \citenamefont {Marcus},\ and\ \citenamefont
  {Aspuru-Guzik}}]{PachonMarcusAspuru-Guzik2015}%
  \BibitemOpen
  \bibfield  {author} {\bibinfo {author} {\bibfnamefont {Leonardo~A.}\
  \bibnamefont {Pach{\'o}n}}, \bibinfo {author} {\bibfnamefont {Andrew~H.}\
  \bibnamefont {Marcus}}, \ and\ \bibinfo {author} {\bibfnamefont {Al{\'a}n}\
  \bibnamefont {Aspuru-Guzik}},\ }\bibfield  {title} {\enquote {\bibinfo
  {title} {Quantum process tomography by 2d fluorescence spectroscopy},}\
  }\href {\doibase 10.1063/1.4919954} {\bibfield  {journal} {\bibinfo
  {journal} {The Journal of Chemical Physics}\ }\textbf {\bibinfo {volume}
  {142}},\ \bibinfo {pages} {212442} (\bibinfo {year} {2015})}\BibitemShut
  {NoStop}%
\bibitem [{\citenamefont {Bialczak}\ \emph {et~al.}(2010)\citenamefont
  {Bialczak}, \citenamefont {Ansmann}, \citenamefont {Hofheinz}, \citenamefont
  {Lucero}, \citenamefont {Neeley}, \citenamefont {O'Connell}, \citenamefont
  {Sank}, \citenamefont {Wang}, \citenamefont {Wenner}, \citenamefont
  {Steffen}, \citenamefont {Cleland},\ and\ \citenamefont
  {Martinis}}]{BialczakAnsmannHofheinz2010}%
  \BibitemOpen
  \bibfield  {author} {\bibinfo {author} {\bibfnamefont {R.~C.}\ \bibnamefont
  {Bialczak}}, \bibinfo {author} {\bibfnamefont {M.}~\bibnamefont {Ansmann}},
  \bibinfo {author} {\bibfnamefont {M.}~\bibnamefont {Hofheinz}}, \bibinfo
  {author} {\bibfnamefont {E.}~\bibnamefont {Lucero}}, \bibinfo {author}
  {\bibfnamefont {M.}~\bibnamefont {Neeley}}, \bibinfo {author} {\bibfnamefont
  {A.~D.}\ \bibnamefont {O'Connell}}, \bibinfo {author} {\bibfnamefont
  {D.}~\bibnamefont {Sank}}, \bibinfo {author} {\bibfnamefont {H.}~\bibnamefont
  {Wang}}, \bibinfo {author} {\bibfnamefont {J.}~\bibnamefont {Wenner}},
  \bibinfo {author} {\bibfnamefont {M.}~\bibnamefont {Steffen}}, \bibinfo
  {author} {\bibfnamefont {A.~N.}\ \bibnamefont {Cleland}}, \ and\ \bibinfo
  {author} {\bibfnamefont {J.~M.}\ \bibnamefont {Martinis}},\ }\bibfield
  {title} {\enquote {\bibinfo {title} {Quantum process tomography of a
  universal entangling gate implemented with {Josephson} phase qubits},}\
  }\href {http://dx.doi.org/10.1038/nphys1639} {\bibfield  {journal} {\bibinfo
  {journal} {Nature Physics}\ }\textbf {\bibinfo {volume} {6}},\ \bibinfo
  {pages} {409} (\bibinfo {year} {2010})}\BibitemShut {NoStop}%
\bibitem [{\citenamefont {Howard}\ \emph {et~al.}(2006)\citenamefont {Howard},
  \citenamefont {Twamley}, \citenamefont {Wittmann}, \citenamefont {Gaebel},
  \citenamefont {Jelezko},\ and\ \citenamefont
  {Wrachtrup}}]{HowardTwamleyWittmann2006}%
  \BibitemOpen
  \bibfield  {author} {\bibinfo {author} {\bibfnamefont {M}~\bibnamefont
  {Howard}}, \bibinfo {author} {\bibfnamefont {J}~\bibnamefont {Twamley}},
  \bibinfo {author} {\bibfnamefont {C}~\bibnamefont {Wittmann}}, \bibinfo
  {author} {\bibfnamefont {T}~\bibnamefont {Gaebel}}, \bibinfo {author}
  {\bibfnamefont {F}~\bibnamefont {Jelezko}}, \ and\ \bibinfo {author}
  {\bibfnamefont {J}~\bibnamefont {Wrachtrup}},\ }\bibfield  {title} {\enquote
  {\bibinfo {title} {Quantum process tomography and {Linblad} estimation of a
  solid-state qubit},}\ }\href {http://stacks.iop.org/1367-2630/8/i=3/a=033}
  {\bibfield  {journal} {\bibinfo  {journal} {New Journal of Physics}\ }\textbf
  {\bibinfo {volume} {8}},\ \bibinfo {pages} {33} (\bibinfo {year}
  {2006})}\BibitemShut {NoStop}%
\bibitem [{\citenamefont {Shabani}\ \emph {et~al.}(2011)\citenamefont
  {Shabani}, \citenamefont {Kosut}, \citenamefont {Mohseni}, \citenamefont
  {Rabitz}, \citenamefont {Broome}, \citenamefont {Almeida}, \citenamefont
  {Fedrizzi},\ and\ \citenamefont {White}}]{ShabaniKosutMohseni2011}%
  \BibitemOpen
  \bibfield  {author} {\bibinfo {author} {\bibfnamefont {A.}~\bibnamefont
  {Shabani}}, \bibinfo {author} {\bibfnamefont {R.~L.}\ \bibnamefont {Kosut}},
  \bibinfo {author} {\bibfnamefont {M.}~\bibnamefont {Mohseni}}, \bibinfo
  {author} {\bibfnamefont {H.}~\bibnamefont {Rabitz}}, \bibinfo {author}
  {\bibfnamefont {M.~A.}\ \bibnamefont {Broome}}, \bibinfo {author}
  {\bibfnamefont {M.~P.}\ \bibnamefont {Almeida}}, \bibinfo {author}
  {\bibfnamefont {A.}~\bibnamefont {Fedrizzi}}, \ and\ \bibinfo {author}
  {\bibfnamefont {A.~G.}\ \bibnamefont {White}},\ }\bibfield  {title} {\enquote
  {\bibinfo {title} {Efficient measurement of quantum dynamics via compressive
  sensing},}\ }\href {\doibase 10.1103/PhysRevLett.106.100401} {\bibfield
  {journal} {\bibinfo  {journal} {Phys. Rev. Lett.}\ }\textbf {\bibinfo
  {volume} {106}},\ \bibinfo {pages} {100401} (\bibinfo {year}
  {2011})}\BibitemShut {NoStop}%
\bibitem [{\citenamefont {Wallman}\ and\ \citenamefont
  {Flammia}(2014)}]{WallmanFlammia2014}%
  \BibitemOpen
  \bibfield  {author} {\bibinfo {author} {\bibfnamefont {Joel~J}\ \bibnamefont
  {Wallman}}\ and\ \bibinfo {author} {\bibfnamefont {Steven~T}\ \bibnamefont
  {Flammia}},\ }\bibfield  {title} {\enquote {\bibinfo {title} {Randomized
  benchmarking with confidence},}\ }\href
  {http://stacks.iop.org/1367-2630/16/i=10/a=103032} {\bibfield  {journal}
  {\bibinfo  {journal} {New Journal of Physics}\ }\textbf {\bibinfo {volume}
  {16}},\ \bibinfo {pages} {103032} (\bibinfo {year} {2014})}\BibitemShut
  {NoStop}%
\bibitem [{\citenamefont {Sanders}\ \emph {et~al.}(2016)\citenamefont
  {Sanders}, \citenamefont {Wallman},\ and\ \citenamefont
  {Sanders}}]{SandersWallmanSanders2016}%
  \BibitemOpen
  \bibfield  {author} {\bibinfo {author} {\bibfnamefont {Yuval~R}\ \bibnamefont
  {Sanders}}, \bibinfo {author} {\bibfnamefont {Joel~J}\ \bibnamefont
  {Wallman}}, \ and\ \bibinfo {author} {\bibfnamefont {Barry~C}\ \bibnamefont
  {Sanders}},\ }\bibfield  {title} {\enquote {\bibinfo {title} {Bounding
  quantum gate error rate based on reported average fidelity},}\ }\href
  {http://stacks.iop.org/1367-2630/18/i=1/a=012002} {\bibfield  {journal}
  {\bibinfo  {journal} {New Journal of Physics}\ }\textbf {\bibinfo {volume}
  {18}},\ \bibinfo {pages} {012002} (\bibinfo {year} {2016})}\BibitemShut
  {NoStop}%
\bibitem [{\citenamefont {Blume-Kohout}\ \emph {et~al.}(2017)\citenamefont
  {Blume-Kohout}, \citenamefont {Gamble}, \citenamefont {Nielsen},
  \citenamefont {Rudinger}, \citenamefont {Mizrahi}, \citenamefont {Fortier},\
  and\ \citenamefont {Maunz}}]{Blume-KohoutGambleNielsen2017}%
  \BibitemOpen
  \bibfield  {author} {\bibinfo {author} {\bibfnamefont {Robin}\ \bibnamefont
  {Blume-Kohout}}, \bibinfo {author} {\bibfnamefont {John~King}\ \bibnamefont
  {Gamble}}, \bibinfo {author} {\bibfnamefont {Erik}\ \bibnamefont {Nielsen}},
  \bibinfo {author} {\bibfnamefont {Kenneth}\ \bibnamefont {Rudinger}},
  \bibinfo {author} {\bibfnamefont {Jonathan}\ \bibnamefont {Mizrahi}},
  \bibinfo {author} {\bibfnamefont {Kevin}\ \bibnamefont {Fortier}}, \ and\
  \bibinfo {author} {\bibfnamefont {Peter}\ \bibnamefont {Maunz}},\ }\bibfield
  {title} {\enquote {\bibinfo {title} {Demonstration of qubit operations below
  a rigorous fault tolerance threshold with gate set tomography},}\ }\href
  {http://dx.doi.org/10.1038/ncomms14485} {\bibfield  {journal} {\bibinfo
  {journal} {Nature Communications}\ }\textbf {\bibinfo {volume} {8}},\
  \bibinfo {pages} {14485} (\bibinfo {year} {2017})}\BibitemShut {NoStop}%
\bibitem [{\citenamefont {Daniels}(1961)}]{Daniels1961}%
  \BibitemOpen
  \bibfield  {author} {\bibinfo {author} {\bibfnamefont {H.~E.}\ \bibnamefont
  {Daniels}},\ }\bibfield  {title} {\enquote {\bibinfo {title} {The asymptotic
  efficiency of a maximum likelihood estimator},}\ }in\ \href
  {https://projecteuclid.org/euclid.bsmsp/1200512164} {\emph {\bibinfo
  {booktitle} {Proc. Fourth Berkeley Symp. on Math. Statist. and Prob., Vol.
  1}}}\ (\bibinfo  {publisher} {University of California Press},\ \bibinfo
  {address} {Berkeley, Calif.},\ \bibinfo {year} {1961})\ pp.\ \bibinfo {pages}
  {151--163}\BibitemShut {NoStop}%
\bibitem [{\citenamefont {Kok}\ and\ \citenamefont
  {Lovett}(2010)}]{KokLovett2010}%
  \BibitemOpen
  \bibfield  {author} {\bibinfo {author} {\bibfnamefont {P.}~\bibnamefont
  {Kok}}\ and\ \bibinfo {author} {\bibfnamefont {B.W.}\ \bibnamefont
  {Lovett}},\ }\href {http://books.google.co.uk/books?id=G2zKNooOeKcC} {\emph
  {\bibinfo {title} {Introduction to Optical Quantum Information Processing}}}\
  (\bibinfo  {publisher} {Cambridge University Press},\ \bibinfo {year}
  {2010})\BibitemShut {NoStop}%
\bibitem [{\citenamefont {Sugiyama}\ \emph {et~al.}(2013)\citenamefont
  {Sugiyama}, \citenamefont {Turner},\ and\ \citenamefont
  {Murao}}]{SugiyamaTurnerMurao2013}%
  \BibitemOpen
  \bibfield  {author} {\bibinfo {author} {\bibfnamefont {Takanori}\
  \bibnamefont {Sugiyama}}, \bibinfo {author} {\bibfnamefont {Peter~S.}\
  \bibnamefont {Turner}}, \ and\ \bibinfo {author} {\bibfnamefont {Mio}\
  \bibnamefont {Murao}},\ }\bibfield  {title} {\enquote {\bibinfo {title}
  {Precision-guaranteed quantum tomography},}\ }\href {\doibase
  10.1103/PhysRevLett.111.160406} {\bibfield  {journal} {\bibinfo  {journal}
  {Phys. Rev. Lett.}\ }\textbf {\bibinfo {volume} {111}},\ \bibinfo {pages}
  {160406} (\bibinfo {year} {2013})}\BibitemShut {NoStop}%
\bibitem [{\citenamefont {Smolin}\ \emph {et~al.}(2012)\citenamefont {Smolin},
  \citenamefont {Gambetta},\ and\ \citenamefont
  {Smith}}]{SmolinGambettaSmith2012}%
  \BibitemOpen
  \bibfield  {author} {\bibinfo {author} {\bibfnamefont {John~A.}\ \bibnamefont
  {Smolin}}, \bibinfo {author} {\bibfnamefont {Jay~M.}\ \bibnamefont
  {Gambetta}}, \ and\ \bibinfo {author} {\bibfnamefont {Graeme}\ \bibnamefont
  {Smith}},\ }\bibfield  {title} {\enquote {\bibinfo {title} {Efficient method
  for computing the maximum-likelihood quantum state from measurements with
  additive {Gaussian} noise},}\ }\href {\doibase
  10.1103/PhysRevLett.108.070502} {\bibfield  {journal} {\bibinfo  {journal}
  {Phys. Rev. Lett.}\ }\textbf {\bibinfo {volume} {108}},\ \bibinfo {pages}
  {070502} (\bibinfo {year} {2012})}\BibitemShut {NoStop}%
\bibitem [{\citenamefont {Teo}(2015)}]{Teo2015}%
  \BibitemOpen
  \bibfield  {author} {\bibinfo {author} {\bibfnamefont {Y.S.}\ \bibnamefont
  {Teo}},\ }\href {https://books.google.co.uk/books?id=yOSiCgAAQBAJ} {\emph
  {\bibinfo {title} {Introduction to Quantum-State Estimation}}}\ (\bibinfo
  {publisher} {World Scientific Publishing Company},\ \bibinfo {year}
  {2015})\BibitemShut {NoStop}%
\bibitem [{\citenamefont {Chuang}\ and\ \citenamefont
  {Nielsen}(1997)}]{ChuangNielsen1997}%
  \BibitemOpen
  \bibfield  {author} {\bibinfo {author} {\bibfnamefont {Isaac~L.}\
  \bibnamefont {Chuang}}\ and\ \bibinfo {author} {\bibfnamefont {M.~A.}\
  \bibnamefont {Nielsen}},\ }\bibfield  {title} {\enquote {\bibinfo {title}
  {Prescription for experimental determination of the dynamics of a quantum
  black box},}\ }\href {\doibase 10.1080/09500349708231894} {\bibfield
  {journal} {\bibinfo  {journal} {Journal of Modern Optics}\ }\textbf {\bibinfo
  {volume} {44}},\ \bibinfo {pages} {2455--2467} (\bibinfo {year}
  {1997})}\BibitemShut {NoStop}%
\bibitem [{\citenamefont {Leung}(2003)}]{Leung2003}%
  \BibitemOpen
  \bibfield  {author} {\bibinfo {author} {\bibfnamefont {Debbie~W.}\
  \bibnamefont {Leung}},\ }\bibfield  {title} {\enquote {\bibinfo {title}
  {Choi's proof as a recipe for quantum process tomography},}\ }\href {\doibase
  10.1063/1.1518554} {\bibfield  {journal} {\bibinfo  {journal} {Journal of
  Mathematical Physics}\ }\textbf {\bibinfo {volume} {44}},\ \bibinfo {pages}
  {528--533} (\bibinfo {year} {2003})}\BibitemShut {NoStop}%
\bibitem [{\citenamefont {Altepeter}\ \emph {et~al.}(2003)\citenamefont
  {Altepeter}, \citenamefont {Branning}, \citenamefont {Jeffrey}, \citenamefont
  {Wei}, \citenamefont {Kwiat}, \citenamefont {Thew}, \citenamefont {O'Brien},
  \citenamefont {Nielsen},\ and\ \citenamefont
  {White}}]{AltepeterBranningJeffrey2003}%
  \BibitemOpen
  \bibfield  {author} {\bibinfo {author} {\bibfnamefont {J.~B.}\ \bibnamefont
  {Altepeter}}, \bibinfo {author} {\bibfnamefont {D.}~\bibnamefont {Branning}},
  \bibinfo {author} {\bibfnamefont {E.}~\bibnamefont {Jeffrey}}, \bibinfo
  {author} {\bibfnamefont {T.~C.}\ \bibnamefont {Wei}}, \bibinfo {author}
  {\bibfnamefont {P.~G.}\ \bibnamefont {Kwiat}}, \bibinfo {author}
  {\bibfnamefont {R.~T.}\ \bibnamefont {Thew}}, \bibinfo {author}
  {\bibfnamefont {J.~L.}\ \bibnamefont {O'Brien}}, \bibinfo {author}
  {\bibfnamefont {M.~A.}\ \bibnamefont {Nielsen}}, \ and\ \bibinfo {author}
  {\bibfnamefont {A.~G.}\ \bibnamefont {White}},\ }\bibfield  {title} {\enquote
  {\bibinfo {title} {Ancilla-assisted quantum process tomography},}\ }\href
  {\doibase 10.1103/PhysRevLett.90.193601} {\bibfield  {journal} {\bibinfo
  {journal} {Phys. Rev. Lett.}\ }\textbf {\bibinfo {volume} {90}},\ \bibinfo
  {pages} {193601} (\bibinfo {year} {2003})}\BibitemShut {NoStop}%
\bibitem [{\citenamefont {Choi}(1975)}]{Choi1975}%
  \BibitemOpen
  \bibfield  {author} {\bibinfo {author} {\bibfnamefont {Man-Duen}\
  \bibnamefont {Choi}},\ }\bibfield  {title} {\enquote {\bibinfo {title}
  {Completely positive linear maps on complex matrices},}\ }\href {\doibase
  https://doi.org/10.1016/0024-3795(75)90075-0} {\bibfield  {journal} {\bibinfo
   {journal} {Linear Algebra and its Applications}\ }\textbf {\bibinfo {volume}
  {10}},\ \bibinfo {pages} {285 -- 290} (\bibinfo {year} {1975})}\BibitemShut
  {NoStop}%
\bibitem [{Note1()}]{Note1}%
  \BibitemOpen
  \bibinfo {note} {A positive semidefinite matrix is denoted by the expression
  $C\succeq 0$, meaning that $C$ belongs to the subset of Hermitian matrices
  with nonnegative eigenvalues.}\BibitemShut {Stop}%
\bibitem [{\citenamefont {Boyd}\ and\ \citenamefont
  {Vandenberghe}(2009)}]{BoydVandenberghe2009}%
  \BibitemOpen
  \bibfield  {author} {\bibinfo {author} {\bibfnamefont {Stephen}\ \bibnamefont
  {Boyd}}\ and\ \bibinfo {author} {\bibfnamefont {Lieven}\ \bibnamefont
  {Vandenberghe}},\ }\href@noop {} {\emph {\bibinfo {title} {Convex
  optimization}}}\ (\bibinfo  {publisher} {Cambridge university press},\
  \bibinfo {year} {2009})\BibitemShut {NoStop}%
\bibitem [{\citenamefont {Plesn{\'\i}k}(2007)}]{Plesnik2007}%
  \BibitemOpen
  \bibfield  {author} {\bibinfo {author} {\bibfnamefont {J{\'a}n}\ \bibnamefont
  {Plesn{\'\i}k}},\ }\bibfield  {title} {\enquote {\bibinfo {title} {Finding
  the orthogonal projection of a point onto an affine subspace},}\ }\href
  {\doibase https://doi.org/10.1016/j.laa.2006.11.003} {\bibfield  {journal}
  {\bibinfo  {journal} {Linear Algebra and its Applications}\ }\textbf
  {\bibinfo {volume} {422}},\ \bibinfo {pages} {455 -- 470} (\bibinfo {year}
  {2007})}\BibitemShut {NoStop}%
\bibitem [{\citenamefont {Bolduc}\ \emph {et~al.}(2017)\citenamefont {Bolduc},
  \citenamefont {Knee}, \citenamefont {Gauger},\ and\ \citenamefont
  {Leach}}]{BolducKneeGauger2017}%
  \BibitemOpen
  \bibfield  {author} {\bibinfo {author} {\bibfnamefont {Eliot}\ \bibnamefont
  {Bolduc}}, \bibinfo {author} {\bibfnamefont {George~C.}\ \bibnamefont
  {Knee}}, \bibinfo {author} {\bibfnamefont {Erik~M.}\ \bibnamefont {Gauger}},
  \ and\ \bibinfo {author} {\bibfnamefont {Jonathan}\ \bibnamefont {Leach}},\
  }\bibfield  {title} {\enquote {\bibinfo {title} {Projected gradient descent
  algorithms for quantum state tomography},}\ }\href {\doibase
  10.1038/s41534-017-0043-1} {\bibfield  {journal} {\bibinfo  {journal} {npj
  Quantum Information}\ }\textbf {\bibinfo {volume} {3}},\ \bibinfo {pages}
  {44} (\bibinfo {year} {2017})}\BibitemShut {NoStop}%
\bibitem [{\citenamefont {Escalante}\ and\ \citenamefont
  {Raydan}(2011)}]{EscalanteRaydan2011}%
  \BibitemOpen
  \bibfield  {author} {\bibinfo {author} {\bibfnamefont {R.}~\bibnamefont
  {Escalante}}\ and\ \bibinfo {author} {\bibfnamefont {M.}~\bibnamefont
  {Raydan}},\ }\href {\doibase 10.1137/9781611971941} {\emph {\bibinfo {title}
  {Alternating Projection Methods}}}\ (\bibinfo  {publisher} {Society for
  Industrial and Applied Mathematics},\ \bibinfo {address} {Philadelphia, PA},\
  \bibinfo {year} {2011})\BibitemShut {NoStop}%
\bibitem [{\citenamefont {Birgin}\ and\ \citenamefont
  {Raydan}(2009)}]{Birgin2009}%
  \BibitemOpen
  \bibfield  {author} {\bibinfo {author} {\bibfnamefont {Ernesto~G.}\
  \bibnamefont {Birgin}}\ and\ \bibinfo {author} {\bibfnamefont {Marcos}\
  \bibnamefont {Raydan}},\ }\enquote {\bibinfo {title} {Dykstra's algorithm and
  robust stopping criteria},}\ in\ \href {\doibase
  10.1007/978-0-387-74759-0_143} {\emph {\bibinfo {booktitle} {Encyclopedia of
  Optimization}}},\ \bibinfo {editor} {edited by\ \bibinfo {editor}
  {\bibfnamefont {Christodoulos~A.}\ \bibnamefont {Floudas}}\ and\ \bibinfo
  {editor} {\bibfnamefont {Panos~M.}\ \bibnamefont {Pardalos}}}\ (\bibinfo
  {publisher} {Springer US},\ \bibinfo {address} {Boston, MA},\ \bibinfo {year}
  {2009})\ pp.\ \bibinfo {pages} {828--833}\BibitemShut {NoStop}%
\bibitem [{\citenamefont {Drusvyatskiy}\ \emph {et~al.}(2015)\citenamefont
  {Drusvyatskiy}, \citenamefont {Li}, \citenamefont {Pelejo}, \citenamefont
  {Voronin},\ and\ \citenamefont {Wolkowicz}}]{DrusvyatskiyLiPelejo2015}%
  \BibitemOpen
  \bibfield  {author} {\bibinfo {author} {\bibfnamefont {Dmitriy}\ \bibnamefont
  {Drusvyatskiy}}, \bibinfo {author} {\bibfnamefont {Chi-Kwong}\ \bibnamefont
  {Li}}, \bibinfo {author} {\bibfnamefont {Diane~Christine}\ \bibnamefont
  {Pelejo}}, \bibinfo {author} {\bibfnamefont {Yuen-Lam}\ \bibnamefont
  {Voronin}}, \ and\ \bibinfo {author} {\bibfnamefont {Henry}\ \bibnamefont
  {Wolkowicz}},\ }\bibfield  {title} {\enquote {\bibinfo {title} {Projection
  methods for quantum channel construction},}\ }\href {\doibase
  10.1007/s11128-015-1024-y} {\bibfield  {journal} {\bibinfo  {journal}
  {Quantum Information Processing}\ }\textbf {\bibinfo {volume} {14}},\
  \bibinfo {pages} {3075--3096} (\bibinfo {year} {2015})}\BibitemShut {NoStop}%
\bibitem [{\citenamefont {Gon{\c c}alves}\ \emph {et~al.}(2016)\citenamefont
  {Gon{\c c}alves}, \citenamefont {Gomes-Ruggiero},\ and\ \citenamefont
  {Lavor}}]{GoncalvesGomes-RuggieroLavor2016}%
  \BibitemOpen
  \bibfield  {author} {\bibinfo {author} {\bibfnamefont {D.S.}\ \bibnamefont
  {Gon{\c c}alves}}, \bibinfo {author} {\bibfnamefont {M.A.}\ \bibnamefont
  {Gomes-Ruggiero}}, \ and\ \bibinfo {author} {\bibfnamefont {C.}~\bibnamefont
  {Lavor}},\ }\bibfield  {title} {\enquote {\bibinfo {title} {A projected
  gradient method for optimization over density matrices},}\ }\href {\doibase
  10.1080/10556788.2015.1082105} {\bibfield  {journal} {\bibinfo  {journal}
  {Optimization Methods and Software}\ }\textbf {\bibinfo {volume} {31}},\
  \bibinfo {pages} {328--341} (\bibinfo {year} {2016})}\BibitemShut {NoStop}%
\bibitem [{\citenamefont {Shang}\ \emph {et~al.}(2017)\citenamefont {Shang},
  \citenamefont {Zhang},\ and\ \citenamefont {Ng}}]{ShangZhangNg2017}%
  \BibitemOpen
  \bibfield  {author} {\bibinfo {author} {\bibfnamefont {Jiangwei}\
  \bibnamefont {Shang}}, \bibinfo {author} {\bibfnamefont {Zhengyun}\
  \bibnamefont {Zhang}}, \ and\ \bibinfo {author} {\bibfnamefont {Hui~Khoon}\
  \bibnamefont {Ng}},\ }\bibfield  {title} {\enquote {\bibinfo {title}
  {Superfast maximum-likelihood reconstruction for quantum tomography},}\
  }\href {\doibase 10.1103/PhysRevA.95.062336} {\bibfield  {journal} {\bibinfo
  {journal} {Phys. Rev. A}\ }\textbf {\bibinfo {volume} {95}},\ \bibinfo
  {pages} {062336} (\bibinfo {year} {2017})}\BibitemShut {NoStop}%
\bibitem [{\citenamefont {Henrion}\ and\ \citenamefont
  {Malick}(2012)}]{HenrionMalick2012}%
  \BibitemOpen
  \bibfield  {author} {\bibinfo {author} {\bibfnamefont {Didier}\ \bibnamefont
  {Henrion}}\ and\ \bibinfo {author} {\bibfnamefont {J{\'e}r{\^o}me}\
  \bibnamefont {Malick}},\ }\enquote {\bibinfo {title} {Projection methods in
  conic optimization},}\ in\ \href {\doibase 10.1007/978-1-4614-0769-0_20}
  {\emph {\bibinfo {booktitle} {Handbook on Semidefinite, Conic and Polynomial
  Optimization}}},\ \bibinfo {editor} {edited by\ \bibinfo {editor}
  {\bibfnamefont {Miguel~F.}\ \bibnamefont {Anjos}}\ and\ \bibinfo {editor}
  {\bibfnamefont {Jean~B.}\ \bibnamefont {Lasserre}}}\ (\bibinfo  {publisher}
  {Springer US},\ \bibinfo {address} {Boston, MA},\ \bibinfo {year} {2012})\
  pp.\ \bibinfo {pages} {565--600}\BibitemShut {NoStop}%
\bibitem [{\citenamefont {Armijo}(1966)}]{Armijo1966}%
  \BibitemOpen
  \bibfield  {author} {\bibinfo {author} {\bibfnamefont {Larry}\ \bibnamefont
  {Armijo}},\ }\bibfield  {title} {\enquote {\bibinfo {title} {Minimization of
  functions having lipschitz continuous first partial derivatives.}}\ }\href
  {https://projecteuclid.org:443/euclid.pjm/1102995080} {\bibfield  {journal}
  {\bibinfo  {journal} {Pacific J. Math.}\ }\textbf {\bibinfo {volume} {16}},\
  \bibinfo {pages} {1--3} (\bibinfo {year} {1966})}\BibitemShut {NoStop}%
\bibitem [{\citenamefont {Bertsekas}(1976)}]{Bertsekas1976}%
  \BibitemOpen
  \bibfield  {author} {\bibinfo {author} {\bibfnamefont {D.}~\bibnamefont
  {Bertsekas}},\ }\bibfield  {title} {\enquote {\bibinfo {title} {On the
  {Goldstein-Levitin-Polyak} gradient projection method},}\ }\href {\doibase
  10.1109/TAC.1976.1101194} {\bibfield  {journal} {\bibinfo  {journal} {IEEE
  Transactions on Automatic Control}\ }\textbf {\bibinfo {volume} {21}},\
  \bibinfo {pages} {174--184} (\bibinfo {year} {1976})}\BibitemShut {NoStop}%
\bibitem [{\citenamefont {Li}\ and\ \citenamefont
  {Cevher}(2017)}]{LiCevher2017}%
  \BibitemOpen
  \bibfield  {author} {\bibinfo {author} {\bibfnamefont {Yen-Huan}\
  \bibnamefont {Li}}\ and\ \bibinfo {author} {\bibfnamefont {Volkan}\
  \bibnamefont {Cevher}},\ }\href@noop {} {\enquote {\bibinfo {title} {Private
  communication},}\ } (\bibinfo {year} {2017})\BibitemShut {NoStop}%
\bibitem [{\citenamefont {Beck}(2014)}]{Beck2014}%
  \BibitemOpen
  \bibfield  {author} {\bibinfo {author} {\bibfnamefont {A.}~\bibnamefont
  {Beck}},\ }\href {https://books.google.co.uk/books?id=VvYeBQAAQBAJ} {\emph
  {\bibinfo {title} {Introduction to Nonlinear Optimization: Theory,
  Algorithms, and Applications with MATLAB}}},\ MOS-SIAM Series on
  Optimization\ (\bibinfo  {publisher} {Society for Industrial and Applied
  Mathematics},\ \bibinfo {year} {2014})\BibitemShut {NoStop}%
\bibitem [{\citenamefont {Kaznady}\ and\ \citenamefont
  {James}(2009)}]{KaznadyJames2009}%
  \BibitemOpen
  \bibfield  {author} {\bibinfo {author} {\bibfnamefont {Max~S.}\ \bibnamefont
  {Kaznady}}\ and\ \bibinfo {author} {\bibfnamefont {Daniel F.~V.}\
  \bibnamefont {James}},\ }\bibfield  {title} {\enquote {\bibinfo {title}
  {Numerical strategies for quantum tomography: Alternatives to full
  optimization},}\ }\href {\doibase 10.1103/PhysRevA.79.022109} {\bibfield
  {journal} {\bibinfo  {journal} {Phys. Rev. A}\ }\textbf {\bibinfo {volume}
  {79}},\ \bibinfo {pages} {022109} (\bibinfo {year} {2009})}\BibitemShut
  {NoStop}%
\bibitem [{\citenamefont {Renes}\ \emph {et~al.}(2004)\citenamefont {Renes},
  \citenamefont {Blume-Kohout}, \citenamefont {Scott},\ and\ \citenamefont
  {Caves}}]{RenesBlume-KohoutScott2004}%
  \BibitemOpen
  \bibfield  {author} {\bibinfo {author} {\bibfnamefont {Joseph~M.}\
  \bibnamefont {Renes}}, \bibinfo {author} {\bibfnamefont {Robin}\ \bibnamefont
  {Blume-Kohout}}, \bibinfo {author} {\bibfnamefont {A.~J.}\ \bibnamefont
  {Scott}}, \ and\ \bibinfo {author} {\bibfnamefont {Carlton~M.}\ \bibnamefont
  {Caves}},\ }\bibfield  {title} {\enquote {\bibinfo {title} {Symmetric
  informationally complete quantum measurements},}\ }\href {\doibase
  10.1063/1.1737053} {\bibfield  {journal} {\bibinfo  {journal} {Journal of
  Mathematical Physics}\ }\textbf {\bibinfo {volume} {45}},\ \bibinfo {pages}
  {2171--2180} (\bibinfo {year} {2004})}\BibitemShut {NoStop}%
\bibitem [{\citenamefont {Merkel}\ \emph {et~al.}(2013)\citenamefont {Merkel},
  \citenamefont {Gambetta}, \citenamefont {Smolin}, \citenamefont {Poletto},
  \citenamefont {C\'orcoles}, \citenamefont {Johnson}, \citenamefont {Ryan},\
  and\ \citenamefont {Steffen}}]{MerkelGambettaSmolin2013}%
  \BibitemOpen
  \bibfield  {author} {\bibinfo {author} {\bibfnamefont {Seth~T.}\ \bibnamefont
  {Merkel}}, \bibinfo {author} {\bibfnamefont {Jay~M.}\ \bibnamefont
  {Gambetta}}, \bibinfo {author} {\bibfnamefont {John~A.}\ \bibnamefont
  {Smolin}}, \bibinfo {author} {\bibfnamefont {Stefano}\ \bibnamefont
  {Poletto}}, \bibinfo {author} {\bibfnamefont {Antonio~D.}\ \bibnamefont
  {C\'orcoles}}, \bibinfo {author} {\bibfnamefont {Blake~R.}\ \bibnamefont
  {Johnson}}, \bibinfo {author} {\bibfnamefont {Colm~A.}\ \bibnamefont {Ryan}},
  \ and\ \bibinfo {author} {\bibfnamefont {Matthias}\ \bibnamefont {Steffen}},\
  }\bibfield  {title} {\enquote {\bibinfo {title} {Self-consistent quantum
  process tomography},}\ }\href {\doibase 10.1103/PhysRevA.87.062119}
  {\bibfield  {journal} {\bibinfo  {journal} {Phys. Rev. A}\ }\textbf {\bibinfo
  {volume} {87}},\ \bibinfo {pages} {062119} (\bibinfo {year}
  {2013})}\BibitemShut {NoStop}%
\bibitem [{\citenamefont {{Blume-Kohout}}\ \emph {et~al.}(2013)\citenamefont
  {{Blume-Kohout}}, \citenamefont {{Gamble}}, \citenamefont {{Nielsen}},
  \citenamefont {{Mizrahi}}, \citenamefont {{Sterk}},\ and\ \citenamefont
  {{Maunz}}}]{Blume-KohoutGambleNielsen2013}%
  \BibitemOpen
  \bibfield  {author} {\bibinfo {author} {\bibfnamefont {R.}~\bibnamefont
  {{Blume-Kohout}}}, \bibinfo {author} {\bibfnamefont {J.~K.}\ \bibnamefont
  {{Gamble}}}, \bibinfo {author} {\bibfnamefont {E.}~\bibnamefont {{Nielsen}}},
  \bibinfo {author} {\bibfnamefont {J.}~\bibnamefont {{Mizrahi}}}, \bibinfo
  {author} {\bibfnamefont {J.~D.}\ \bibnamefont {{Sterk}}}, \ and\ \bibinfo
  {author} {\bibfnamefont {P.}~\bibnamefont {{Maunz}}},\ }\href@noop {}
  {\enquote {\bibinfo {title} {{Robust, self-consistent, closed-form tomography
  of quantum logic gates on a trapped ion qubit}},}\ } (\bibinfo {year}
  {2013}),\ \bibinfo {note} {pre-print},\ \Eprint
  {http://arxiv.org/abs/1310.4492} {arXiv:1310.4492 [quant-ph]} \BibitemShut
  {NoStop}%
\bibitem [{\citenamefont {Gilchrist}\ \emph {et~al.}(2005)\citenamefont
  {Gilchrist}, \citenamefont {Langford},\ and\ \citenamefont
  {Nielsen}}]{GilchristLangfordNielsen2005}%
  \BibitemOpen
  \bibfield  {author} {\bibinfo {author} {\bibfnamefont {Alexei}\ \bibnamefont
  {Gilchrist}}, \bibinfo {author} {\bibfnamefont {Nathan~K.}\ \bibnamefont
  {Langford}}, \ and\ \bibinfo {author} {\bibfnamefont {Michael~A.}\
  \bibnamefont {Nielsen}},\ }\bibfield  {title} {\enquote {\bibinfo {title}
  {Distance measures to compare real and ideal quantum processes},}\ }\href
  {\doibase 10.1103/PhysRevA.71.062310} {\bibfield  {journal} {\bibinfo
  {journal} {Phys. Rev. A}\ }\textbf {\bibinfo {volume} {71}},\ \bibinfo
  {pages} {062310} (\bibinfo {year} {2005})}\BibitemShut {NoStop}%
\bibitem [{\citenamefont {Fiur{\'a}{\v s}ek}\ and\ \citenamefont
  {Hradil}(2001)}]{Fiur_ek_2001}%
  \BibitemOpen
  \bibfield  {author} {\bibinfo {author} {\bibfnamefont {Jarom{\'\i}r}\
  \bibnamefont {Fiur{\'a}{\v s}ek}}\ and\ \bibinfo {author} {\bibfnamefont
  {Zden{\v e}k}\ \bibnamefont {Hradil}},\ }\bibfield  {title} {\enquote
  {\bibinfo {title} {Maximum-likelihood estimation of quantum processes},}\
  }\href {\doibase 10.1103/PhysRevA.63.020101} {\bibfield  {journal} {\bibinfo
  {journal} {Phys. Rev. A}\ }\textbf {\bibinfo {volume} {63}},\ \bibinfo
  {pages} {020101} (\bibinfo {year} {2001})}\BibitemShut {NoStop}%
\bibitem [{\citenamefont {{\v R}eh{\'a}{\v c}ek}\ \emph
  {et~al.}(2007)\citenamefont {{\v R}eh{\'a}{\v c}ek}, \citenamefont {Hradil},
  \citenamefont {Knill},\ and\ \citenamefont
  {Lvovsky}}]{RehacekHradilKnill2007}%
  \BibitemOpen
  \bibfield  {author} {\bibinfo {author} {\bibfnamefont {Jaroslav}\
  \bibnamefont {{\v R}eh{\'a}{\v c}ek}}, \bibinfo {author} {\bibfnamefont
  {Zden{\v e}k}\ \bibnamefont {Hradil}}, \bibinfo {author} {\bibfnamefont
  {E.}~\bibnamefont {Knill}}, \ and\ \bibinfo {author} {\bibfnamefont {A.~I.}\
  \bibnamefont {Lvovsky}},\ }\bibfield  {title} {\enquote {\bibinfo {title}
  {Diluted maximum-likelihood algorithm for quantum tomography},}\ }\href
  {\doibase 10.1103/PhysRevA.75.042108} {\bibfield  {journal} {\bibinfo
  {journal} {Phys. Rev. A}\ }\textbf {\bibinfo {volume} {75}},\ \bibinfo
  {pages} {042108} (\bibinfo {year} {2007})}\BibitemShut {NoStop}%
\bibitem [{\citenamefont {Anis}\ and\ \citenamefont
  {Lvovsky}(2012)}]{AnisLvovsky2012}%
  \BibitemOpen
  \bibfield  {author} {\bibinfo {author} {\bibfnamefont {Aamir}\ \bibnamefont
  {Anis}}\ and\ \bibinfo {author} {\bibfnamefont {A~I}\ \bibnamefont
  {Lvovsky}},\ }\bibfield  {title} {\enquote {\bibinfo {title}
  {Maximum-likelihood coherent-state quantum process tomography},}\ }\href
  {http://stacks.iop.org/1367-2630/14/i=10/a=105021} {\bibfield  {journal}
  {\bibinfo  {journal} {New Journal of Physics}\ }\textbf {\bibinfo {volume}
  {14}},\ \bibinfo {pages} {105021} (\bibinfo {year} {2012})}\BibitemShut
  {NoStop}%
\bibitem [{\citenamefont {Cooper}\ \emph {et~al.}(2015)\citenamefont {Cooper},
  \citenamefont {Slade}, \citenamefont {Karpi{\'n}ski},\ and\ \citenamefont
  {Smith}}]{CooperSladeKarpinski2015}%
  \BibitemOpen
  \bibfield  {author} {\bibinfo {author} {\bibfnamefont {Merlin}\ \bibnamefont
  {Cooper}}, \bibinfo {author} {\bibfnamefont {Eirion}\ \bibnamefont {Slade}},
  \bibinfo {author} {\bibfnamefont {Micha{\l}}\ \bibnamefont {Karpi{\'n}ski}},
  \ and\ \bibinfo {author} {\bibfnamefont {Brian~J}\ \bibnamefont {Smith}},\
  }\bibfield  {title} {\enquote {\bibinfo {title} {Characterization of
  conditional state-engineering quantum processes by coherent state quantum
  process tomography},}\ }\href
  {http://stacks.iop.org/1367-2630/17/i=3/a=033041} {\bibfield  {journal}
  {\bibinfo  {journal} {New Journal of Physics}\ }\textbf {\bibinfo {volume}
  {17}},\ \bibinfo {pages} {033041} (\bibinfo {year} {2015})}\BibitemShut
  {NoStop}%
\bibitem [{\citenamefont {Fedorov}\ \emph {et~al.}(2015)\citenamefont
  {Fedorov}, \citenamefont {Fedorov}, \citenamefont {Kurochkin},\ and\
  \citenamefont {Lvovsky}}]{FedorovFedorovKurochkin2015}%
  \BibitemOpen
  \bibfield  {author} {\bibinfo {author} {\bibfnamefont {Ilya~A}\ \bibnamefont
  {Fedorov}}, \bibinfo {author} {\bibfnamefont {Aleksey~K}\ \bibnamefont
  {Fedorov}}, \bibinfo {author} {\bibfnamefont {Yury~V}\ \bibnamefont
  {Kurochkin}}, \ and\ \bibinfo {author} {\bibfnamefont {A~I}\ \bibnamefont
  {Lvovsky}},\ }\bibfield  {title} {\enquote {\bibinfo {title} {Tomography of a
  multimode quantum black box},}\ }\href
  {http://stacks.iop.org/1367-2630/17/i=4/a=043063} {\bibfield  {journal}
  {\bibinfo  {journal} {New Journal of Physics}\ }\textbf {\bibinfo {volume}
  {17}},\ \bibinfo {pages} {043063} (\bibinfo {year} {2015})}\BibitemShut
  {NoStop}%
\bibitem [{\citenamefont {Grant}\ and\ \citenamefont
  {Boyd}(2014)}]{GrantBoyd2014}%
  \BibitemOpen
  \bibfield  {author} {\bibinfo {author} {\bibfnamefont {Michael}\ \bibnamefont
  {Grant}}\ and\ \bibinfo {author} {\bibfnamefont {Stephen}\ \bibnamefont
  {Boyd}},\ }\href@noop {} {\enquote {\bibinfo {title} {{CVX}: Matlab software
  for disciplined convex programming, version 2.1},}\ }\bibinfo {howpublished}
  {\url{http://cvxr.com/cvx}} (\bibinfo {year} {2014})\BibitemShut {NoStop}%
\bibitem [{\citenamefont {Jiang}\ \emph {et~al.}(2013)\citenamefont {Jiang},
  \citenamefont {Luo},\ and\ \citenamefont {Fu}}]{JiangLuoFu2013}%
  \BibitemOpen
  \bibfield  {author} {\bibinfo {author} {\bibfnamefont {Min}\ \bibnamefont
  {Jiang}}, \bibinfo {author} {\bibfnamefont {Shunlong}\ \bibnamefont {Luo}}, \
  and\ \bibinfo {author} {\bibfnamefont {Shuangshuang}\ \bibnamefont {Fu}},\
  }\bibfield  {title} {\enquote {\bibinfo {title} {Channel-state duality},}\
  }\href {\doibase 10.1103/PhysRevA.87.022310} {\bibfield  {journal} {\bibinfo
  {journal} {Phys. Rev. A}\ }\textbf {\bibinfo {volume} {87}},\ \bibinfo
  {pages} {022310} (\bibinfo {year} {2013})}\BibitemShut {NoStop}%
\bibitem [{\citenamefont {Sacchi}(2001)}]{Sacchi2001}%
  \BibitemOpen
  \bibfield  {author} {\bibinfo {author} {\bibfnamefont {Massimiliano~F.}\
  \bibnamefont {Sacchi}},\ }\bibfield  {title} {\enquote {\bibinfo {title}
  {Maximum-likelihood reconstruction of completely positive maps},}\ }\href
  {\doibase 10.1103/PhysRevA.63.054104} {\bibfield  {journal} {\bibinfo
  {journal} {Phys. Rev. A}\ }\textbf {\bibinfo {volume} {63}},\ \bibinfo
  {pages} {054104} (\bibinfo {year} {2001})}\BibitemShut {NoStop}%
\bibitem [{\citenamefont {Fiurasek}\ and\ \citenamefont
  {Hradil}(2001)}]{FiurasekHradil2001}%
  \BibitemOpen
  \bibfield  {author} {\bibinfo {author} {\bibfnamefont {Jaromir}\ \bibnamefont
  {Fiurasek}}\ and\ \bibinfo {author} {\bibfnamefont {Zdenek}\ \bibnamefont
  {Hradil}},\ }\bibfield  {title} {\enquote {\bibinfo {title} {{Comment on
  "Maximum likelihood reconstruction of CP maps", quant-ph/0009104}},}\ }\href
  {http://arxiv.org/abs/quant-ph/0101048v1} {\  (\bibinfo {year} {2001})},\
  \Eprint {http://arxiv.org/abs/quant-ph/0101048v1} {arXiv:quant-ph/0101048v1
  [quant-ph]} \BibitemShut {NoStop}%
\bibitem [{\citenamefont {Bongioanni}\ \emph {et~al.}(2010)\citenamefont
  {Bongioanni}, \citenamefont {Sansoni}, \citenamefont {Sciarrino},
  \citenamefont {Vallone},\ and\ \citenamefont
  {Mataloni}}]{BongioanniSansoniSciarrino2010}%
  \BibitemOpen
  \bibfield  {author} {\bibinfo {author} {\bibfnamefont {Irene}\ \bibnamefont
  {Bongioanni}}, \bibinfo {author} {\bibfnamefont {Linda}\ \bibnamefont
  {Sansoni}}, \bibinfo {author} {\bibfnamefont {Fabio}\ \bibnamefont
  {Sciarrino}}, \bibinfo {author} {\bibfnamefont {Giuseppe}\ \bibnamefont
  {Vallone}}, \ and\ \bibinfo {author} {\bibfnamefont {Paolo}\ \bibnamefont
  {Mataloni}},\ }\bibfield  {title} {\enquote {\bibinfo {title} {Experimental
  quantum process tomography of non-trace-preserving maps},}\ }\href {\doibase
  10.1103/PhysRevA.82.042307} {\bibfield  {journal} {\bibinfo  {journal} {Phys.
  Rev. A}\ }\textbf {\bibinfo {volume} {82}},\ \bibinfo {pages} {042307}
  (\bibinfo {year} {2010})}\BibitemShut {NoStop}%
\bibitem [{\citenamefont {Lewis}\ and\ \citenamefont
  {Malick}(2008)}]{LewisMalick2008}%
  \BibitemOpen
  \bibfield  {author} {\bibinfo {author} {\bibfnamefont {Adrian~S.}\
  \bibnamefont {Lewis}}\ and\ \bibinfo {author} {\bibfnamefont
  {J{\'e}r{\^o}me}\ \bibnamefont {Malick}},\ }\bibfield  {title} {\enquote
  {\bibinfo {title} {Alternating projections on manifolds},}\ }\href {\doibase
  10.1287/moor.1070.0291} {\bibfield  {journal} {\bibinfo  {journal}
  {Mathematics of Operations Research}\ }\textbf {\bibinfo {volume} {33}},\
  \bibinfo {pages} {216--234} (\bibinfo {year} {2008})}\BibitemShut {NoStop}%
\bibitem [{\citenamefont {Bertlmann}\ and\ \citenamefont
  {Krammer}(2008)}]{BertlmannKrammer2008}%
  \BibitemOpen
  \bibfield  {author} {\bibinfo {author} {\bibfnamefont {Reinhold~A}\
  \bibnamefont {Bertlmann}}\ and\ \bibinfo {author} {\bibfnamefont {Philipp}\
  \bibnamefont {Krammer}},\ }\bibfield  {title} {\enquote {\bibinfo {title}
  {Bloch vectors for qudits},}\ }\href
  {http://stacks.iop.org/1751-8121/41/i=23/a=235303} {\bibfield  {journal}
  {\bibinfo  {journal} {Journal of Physics A: Mathematical and Theoretical}\
  }\textbf {\bibinfo {volume} {41}},\ \bibinfo {pages} {235303} (\bibinfo
  {year} {2008})}\BibitemShut {NoStop}%
\bibitem [{\citenamefont {Bruzda}\ \emph {et~al.}(2009)\citenamefont {Bruzda},
  \citenamefont {Cappellini}, \citenamefont {Sommers},\ and\ \citenamefont
  {{\.Z}yczkowski}}]{BruzdaCappelliniSommers2009}%
  \BibitemOpen
  \bibfield  {author} {\bibinfo {author} {\bibfnamefont {Wojciech}\
  \bibnamefont {Bruzda}}, \bibinfo {author} {\bibfnamefont {Valerio}\
  \bibnamefont {Cappellini}}, \bibinfo {author} {\bibfnamefont
  {Hans-J{\"u}rgen}\ \bibnamefont {Sommers}}, \ and\ \bibinfo {author}
  {\bibfnamefont {Karol}\ \bibnamefont {{\.Z}yczkowski}},\ }\bibfield  {title}
  {\enquote {\bibinfo {title} {Random quantum operations},}\ }\href {\doibase
  https://doi.org/10.1016/j.physleta.2008.11.043} {\bibfield  {journal}
  {\bibinfo  {journal} {Physics Letters A}\ }\textbf {\bibinfo {volume}
  {373}},\ \bibinfo {pages} {320 -- 324} (\bibinfo {year} {2009})}\BibitemShut
  {NoStop}%
\end{thebibliography}%
\end{document}